\documentclass[11pt]{article}
\pdfoutput=1
\usepackage{amsmath}
\usepackage{amssymb}
\usepackage{amsthm}
\usepackage{color}
\usepackage{hyperref}
\usepackage{mathtools}
\usepackage[toc,page]{appendix}
\usepackage[lofdepth,lotdepth]{subfig}
\usepackage{graphicx}
\usepackage{authblk}

\usepackage{epsf,epsfig,amsmath,amssymb,dsfont}
\usepackage{amsthm,mathrsfs}
\usepackage{graphicx} 
\usepackage{multirow}
\usepackage{float}
\usepackage{colortbl}
\usepackage[table]{xcolor}
\usepackage{varioref}

\usepackage{subfig}

\newcommand{\be}{\begin{equation}}
\newcommand{\ee}{\end{equation}}
\newcommand{\baln}{\begin{align}}
\newcommand{\ealn}{\end{align}}
\newcommand{\ben}{\begin{equation*}}
\newcommand{\een}{\end{equation*}}
\newcommand{\idelta}{i\Delta}
\newcommand{\A}{\mathcal{A}}
\newcommand{\kerd}{\text{ker}{\,\Delta}}
\newcommand{\kerr}{\text{ker}{\,R}}

\long\def\symbolfootnote[#1]#2{\begingroup%
\def\thefootnote{\fnsymbol{footnote}}\footnote[#1]{#2}\endgroup}

\newcommand{\fr}{\frac}

\newcommand\CS{\mathcal{C}}

\newcommand{\cau}{\mathcal{K}}

\title{On the Entanglement Entropy of Quantum Fields in Causal Sets}

\author{Alessio Belenchia$^{1}$, Dionigi M. T. Benincasa$^{2,3,4}$, Marco Letizia$^{3,4}$ and Stefano Liberati$^{3,4}$}

\affil{$^1$Institute for Quantum Optics and Quantum Information (IQOQI), Boltzmanngasse 3 1090 Vienna, Austria}
\affil{$^2$School of Theoretical Physics, Dublin Institute for Advanced Studies, 10 Burlington Road, Dublin 4, Ireland}
\affil{$^3$SISSA - International School for Advanced Studies, Via Bonomea 265, 34136 Trieste, Italy}
\affil{$^4$INFN sezione di Trieste, via Valerio 2, Trieste, Italy}
\date{\today}

\begin{document}

\maketitle

\begin{abstract}

In order to understand the detailed mechanism by which a fundamental discreteness can provide a finite entanglement entropy, we consider the entanglement entropy of two classes of free massless scalar fields on causal sets 
that are well approximated by causal diamonds in Minkowski spacetime of dimensions 2,3 and 4. 
The first class is defined from discretised versions of the continuum retarded Green functions,
while the second uses the causal set's retarded nonlocal d'Alembertians parametrised by a length scale $l_k$. 
In both cases we provide numerical evidence that the area law is recovered when the double-cutoff 
prescription proposed in~\cite{Sorkin:2016pbz} is imposed. 
We discuss in detail the need for this double cutoff by studying the effect of two cutoffs 
on the quantum field and, in particular, on the entanglement entropy, in isolation.
In so doing, we get a novel interpretation for why these two cutoff are necessary, and the different roles 
they play in making the entanglement entropy on causal sets finite.
\end{abstract}

\section{Introduction}\label{introduction}

Of the many hints from general relativity and quantum field theory 
that the 4-dimensional, smooth manifold structure of spacetime may be radically different,
possibly discrete, at the fundamental level, the finiteness 
of the entropy of black holes is the most telling of all. 
Indeed, the Bekenstein--Hawking entropy~\cite{PhysRevD.7.2333} for a black hole of area $A$, $S_{BH} = A/4l_p^2$, 
strongly suggests an explanation in terms of a discrete, Planck-scale, spacetime substructure. 
 
Since this relation was first discovered there have been various proposals attempting to explain what this entropy is, 
and in particular what degrees of freedom it is counting. One of the first  attempts 
was in terms of the entanglement entropy of quantum fields living on 
the black hole spacetime \cite{sorkin1983entropy,PhysRevD.34.373}.

Entanglement entropy is a direct consequence of the correlation structure of states in a local 
relativistic quantum field theory.
Indeed, these states live in a Hilbert space that can be associated to a Cauchy surface of the 
spacetime and, since Cauchy surfaces are achronal, can be decomposed into a tensor product\footnote{Such 
a tensor product decomposition is not always possible. We will see an explicit example of this failure later on in the paper.}
$\mathcal{H}_A\otimes\mathcal{H}_B$, where $\mathcal{H}_{A,B}$
are subspaces of $\mathcal{H}_\Sigma$ associated to disjoint subregions 
$A,B\subset\Sigma$ such that $A\cup B = \Sigma$. 
Therefore, given a state $\rho_{\Sigma}\in \mathcal{H}_\Sigma$, 
one can trace over degrees of freedom associated to a subregion of the surface, say $A$, to form a reduced density 
matrix $\rho_B = \text{Tr}_A(\rho_\Sigma)$. 
If the state $\rho_\Sigma$ is pure, then $\rho_B$ will not be pure in general and as such will have a non-vanishing
von-Neumann entropy associated to it $S_B = -\text{Tr}\rho_B\log \rho_B$.

It turns out that the entanglement entropy of the vacuum state of a quantum field is not only nonzero, 
it is infinite. This divergence can be traced back to the fact that the field lives on a continuum, thus allowing
for correlations in the state over arbitrarily small scales. These infinitely many short range correlations 
get severed by the tracing procedure, thus leading to an infinite entropy of the reduced density matrix.

It is not surprising then, that this divergence can be removed by introducing an ultraviolet cutoff that effectively 
restricts the correlations that are traced over to be on length scales greater than a cutoff scale.
When the field is in its vacuum state this truncation leads to the famous area-law 
\be
S(\rho_B) = \fr{\text{Area}(\partial A)}{l^{d-2}}
\ee
where $\text{Area}(\partial A)$ is the area of the codimension-2 surface separating $A$ and $B$, and $l$ is 
the UV cutoff.\footnote{Note that we generically refer to this as the ``area" law although in spacetime dimensions 
$d\ne4$ it is not really an area in the usual sense of the word.}

As is usually the case when introducing a cutoff that renders an infinite quantity finite, 
this can be thought of as a regulator that effectively fills in for whatever the more 
fundamental, {\it finite}, description of the system may be.
In this paper we consider a particular fundamental model that 
allows for a (in principle) finite account of entanglement 
entropy, namely causal set theory. 

Causal set theory postulates that the fundamental structure of spacetime is a locally
finite partial order, also known as a causal set, or causet for short.\footnote{For modern 
reviews of causal set theory see \cite{Surya:2011yh,Dowker:aza}} 
The order relation of the causal set is taken to 
underly the macroscopic causal ordering of spacetime events, and the postulate of local finiteness
ensures that the structure is truly discrete. We denote the discreteness length scale by $\ell$. 
In a sense, one can think of causal sets as being
made up of ``atoms of spacetime" (the causal set elements) whose relation is one 
of causal antecedence/descendence (the partial order). 

We take here the view that a more fundamental description of quantum field theory on a 
fixed background continuum spacetime is that of a quantum field theory on a fixed,
background causal set. Within this context one might expect that the causal set's discreteness, 
encoded by $\ell$, 
will guarantee a finite entanglement entropy. However,
as we will see shortly, this is not the case in general, albeit the kind of divergences that we will
encounter are of a very different nature to the one that arises in the continuum, 
having more to do with a lack of ``proper" equations of motion,
than the existence of vacuum correlations over arbitrarily small scales. 

It was first noticed in~\cite{Sorkin:2016pbz} that, by introducing two cutoffs in the calculation
of the entanglement entropy for a massless scalar field living on a sprinkling of 
a finite, globally hyperbolic region of 2-dimensional Minkowski spacetime, the area law could
be recovered, together with the well-known coefficient of $1/3$.

In this paper we build on the work of~\cite{Sorkin:2016pbz} by extending their results to higher dimensions
and for a class of quantum field theories parametrised by a nonlocality scale $l_k$. Using their double cutoff
prescription, we provide robust numerical evidence for an area law in 2$d$ for the field theory parametrised by $l_k$, 
and preliminary numerical evidence for an area law in 3 and 4$d$, with important differences
between the theories parametrised by $l_k>\ell$ and those with just the scale $\ell$. In the process we analyse the effect of
the two cutoffs by studying them individually and discussing their origin. 

While one of the cutoffs can be seen as analogous to
the cutoff that renders the entanglement entropy finite in the continuum, and is in fact necessary
in order to make the entropy finite even in the causal set (although for very different reasons), the
other does not have a continuum counterpart, but again appears to be necessary in order to match
the continuum's result in cases where the entropy is known to vanish. Although the impact of both
cutoffs, when considered alone, can be understood to some extent, it is still somewhat of a mystery
why introducing both cutoffs together results in an entanglement entropy that matches that of the continuum.
We discuss potential ways of resolving this puzzle in the final section.

The rest of the paper is organised as follows. In Section \ref{sec:EE-causet} we describe the relevant tools 
necessary for the construction of free, massless, scalar quantum field theories on causal sets and Sorkin's prescription 
for the calculation of their entanglement entropy, which we refer to as the Sorkin entropy from here on.
In Section~\ref{sec:SE} we present the results concerning the computation of the entanglement entropy 
in 2 3 and 4 dimensions with the double cutoff prescription of~\cite{Sorkin:2016pbz}.
In Section~\ref{cutoffs} we deal with the origin and interpretation of such prescription both in the local and 
non-local field theory cases.
Finally in Section~\ref{secdisc} we discuss possible interpretations of our findings and lessons that 
can be learned therefrom.

\section{Entanglement Entropy on Causal Sets}
\label{sec:EE-causet}

A causal set (or causet) is a locally finite partial order, $(\CS,\preceq)$. 
Local finiteness is the condition that the cardinality of any order interval is finite, 
where the (inclusive) order interval between a pair of elements $y\prec x$ is defined to be 
$I(x,y) := \{z\in \CS \,|\, y\preceq z \preceq x\}$. We write $x \prec y$ when $x \preceq y$ and $x \ne y$.

Causal sets corresponding to continuum spacetimes are known as sprinklings. 
More specifically a sprinkling is a kinematical process used to generate a causet from a $d$-dimensional 
spacetime $(\mathcal{M},g)$: it is a Poisson process of selecting points in $\mathcal{M}$ 
with density $\rho$ so that the expected  number of points sprinkled in a region of spacetime volume $V$ is $\rho V$. 
This process thus generates a causet whose elements are the sprinkled points and whose order is that induced by the 
manifold's causal order restricted to the sprinkled points. 
We say that a causet $\CS$ is well approximated by a spacetime $(\mathcal{M}, g)$ if it could have been 
generated by sprinkling into $(\mathcal{M}, g)$, with relatively high probability.

Because sprinkled causets are both discrete and locally Lorentz invariant~\cite{Bombelli:2006nm}, they are also nonlocal,
in the sense that the valency of every element is infinite (or very large when curvature limits
Lorentz symmetry)~\cite{Sorkin:2007qi}. This radical nonlocality appears to be both a feature and a misfeature at the 
same time. On the one hand, it offers hope that a continuum regime of the sum-over-histories dynamics
of causal sets exists~\cite{PhysRevLett.104.181301}, on the other hand, it makes for a tough challenge when trying to define (quasi-)local
concepts. 

A particularly striking example of the latter issue is in the definition of a Cauchy-like surface on a causal set.
Cauchy surfaces are first and foremost achronal sets, and this property is easy to reproduce on a causet 
by selecting a subset of elements that are all unrelated to each other, i.e. an antichain. A maximal antichain
is a an antichain that is not a proper subset of any other antichain, and is the closest analogue of a Cauchy surface
in a causal set. But while Cauchy surfaces in the continuum ``capture" all of the information propagating 
in a spacetime -- and as such provide the structure needed for an initial-value formulation for particles
and fields, including the gravitational one itself -- in the causal set do not 
have this property.\footnote{In fact, even arbitrary large thickenings of maximal antichains will in general
fail to have the property that every causal ``curve" in the causet intersects it exactly once. The only such
``surface" with this property, in general, being the whole causal set itself.}
 
The absence of ``surfaces" on which initial data can be specified means that standard quantisation techniques, 
in which a Hilbert space of states is associated to a moment of time, cannot be applied to
causal sets. Instead one must look for an alternative formulation of quantum field theory that does not rely
on the mathematical machinery available when one has a well-defined Cauchy problem.
In the following section we describe such a formulation.

\subsection{The Sorkin--Johnston Ground State and Causal Set Green Functions}

Unlike traditional approaches to QFT, that start from the equations of motion, 
the Sorkin--Johnston (SJ) prescription starts from the retarded Green function $G_{xy}$. We will not go into the details of the approach here (the interested reader is invited to look at \cite{Sorkin:2017fcp}
for a detailed pedagogical introduction), but will point out that one can schematically depict the logical sequence 
of steps underlying the procedure as
$$G \longrightarrow i\Delta \longrightarrow W,$$
where $i\Delta= [\phi,\phi] = G - G^T$ is the Pauli--Jordan function and $W = \langle\phi\phi \rangle$
is the vacuum Wightman function (the superscript $T$ denotes the transpose).

More specifically the Pauli--Jordan function $\Delta$ 
provides information about $W$ through the relation 
\be
\idelta = W - W^*.
\ee
The SJ-vacuum can then be defined as 
\be
W = \text{Pos}(\idelta) = R + \fr{i}{2}\Delta
\ee
where Pos stands for positive part and $R := \langle \{\phi,\phi\}\rangle = \sqrt{-\Delta^2}/2$. In the literature, $R$ is usually referred to as
the Hadamard two-point function and, in a free QFT, is the part of the Wightman function that gives
information about the state.
This prescription therefore makes it possible 
to define a (distinguished) ground state for any region of spacetime or causal set, 
given a retarded Green function alone. 

Consider now a real, free massless scalar field, $\phi$, living on the causal set.
To be able to exploit the SJ prescription we need a definition for its retarded Green function. 
Thus far in the causal set literature, almost exclusively 
discretised versions of the continuum Green functions have been considered. For example in $d=2,3$ and 4 dimensions 
these are given by
\begin{align}
& G^{(2)}_{xy}=\frac{1}{2}C_{xy}\\
& G^{(3)}_{xy}=\frac{1}{2 \pi }\left(\frac{\pi\rho}{12}\right)^{1/3} \left(\left(C+\mathbb{I}\right)^2\right)_{xy}^{-1/3}\\
& G^{(4)}_{xy}=\frac{\sqrt{\rho}}{2 \pi  \sqrt{6}} L_{xy},
\label{eq:loc}
\end{align}
for $x\prec y$ and zero otherwise, respectively. 
Here $\rho=\ell^{-d}$, $\ell$ is the discreteness scale, the matrix $C_{xy}=1$ if $x\prec y$ and zero otherwise, while $L_{xy}=1$ if $y\prec x$ 
and there are no elements $z\in\CS$ such that $y\prec z\prec x$ and zero otherwise, 
for all $x,y,z\in\mathcal{C}$. It can be shown that, when averaged over all sprinkling of 
Minkowski spacetime of their respective dimensions, these operators reduce to the standard 
continuum retarded Green functions of $\Box$ in the limit $\rho\rightarrow\infty$~\cite{Johnston:2008aa,Johnston:2010aa}.
\footnote{Note that the 2$d$ Green function on the causal set is unique in that its expectation value 
is {\it exactly} the continuum Green function for all values of $\rho$.}

There also exists another prescription for defining Green functions on causal sets 
that makes use of retarded d'Alembertian operators. 
It was shown in \cite{Sorkin:2007qi} that despite the causal set's radical nonlocality, it is still possible to define 
wave operators with the property that when averaged over sprinklings of Minkowski spacetime (i.e.~the 
continuum limit), they give rise to non-local operators that reduce to the standard d'Alembertian 
operator in the local limit $\ell\rightarrow0$. 
Following Sorkin's original paper such operators have been defined in all dimensions 
(and for an arbitrary number of layers, although we will only consider the minimal cases here) 
and are given by 
(see~\cite{Glaser:2013xha} for further details)
\begin{equation}\label{opn}
B^{(d)}_{\rho}\phi(x)=\rho^{2/d}\left(a\,\phi(x)+\sum_{n=0}^{L_{\rm max}}b_{n}\sum_{y\in I_{n}(x)}\phi(y)\right),
\end{equation}
where $d$ labels the spacetime dimension for which the operator is defined,
$a,b_{n}$ are dimension dependent coefficients, 
$I_{n}(x)$ represents the set of past $n$-th nearest neighbours to $x$, and the various
coefficients in dimensions 2,3 and 4 are given by
\begin{table}[h!]
\hspace{-1.77cm}
\label{my-label}
\begin{tabular}{|c|c|c|c|c|c|}
\hline
 & $a$ & $b_{0}$ & $b_{1}$ & $b_{2}$ & $b_{3}$  \\ \hline
$d=2$ & -2 & 4 & -8 & 4 &  \\ \hline
$d=3$ & $-\frac{1}{\Gamma[5/3]}\left(\frac{\pi}{3\sqrt{2}}\right)^{2/3}$ & $\frac{1}{\Gamma[5/3]}\left(\frac{\pi}{3\sqrt{2}}\right)^{2/3}$ & $-\frac{27}{8\Gamma[5/3]}\left(\frac{\pi}{3\sqrt{2}}\right)^{2/3}$ & $\frac{9}{4\Gamma[5/3]}\left(\frac{\pi}{3\sqrt{2}}\right)^{2/3}$ &  \\ \hline
$d=4$ & $-4/\sqrt{6}$ & $4/\sqrt{6}$ & $-36/\sqrt{6}$ & $64/\sqrt{6}$ & $-32/\sqrt{6}$ \\ \hline
\end{tabular}
\caption{Table of coefficients in Eq.\eqref{opn} for $d=2,3,4$. The number of coefficients for every dimension corresponds to the ``minimal'' non-local operators, i.e., the operators constructed with the minimum number of layers.}
\end{table}

The operators~\eqref{nonlocalop} are nonsingular and can therefore be inverted 
to give (nonlocal) retarded Green functions, $G^{(d)} := (B^{(d)})^{-1}$.

In fact, we will be dealing with slightly generalised versions of these operators that are parametrised
by a nonlocality length scale, $l_k\ge\ell$; where $l_k=\ell$
is the minimal amount of nonlocality that one can get away with without spoiling the local limit. 
They are defined as 
\begin{equation}\label{nonlocalop}
    \tilde{B}^{(d)}_{\rho}\phi(x)=(\epsilon\rho)^{2/d}\left(a\,\phi(x)+\sum_{m=0}^{\infty}\tilde{b}_{m}\sum_{y\in I_{m}(x)}\phi(y)\right),
\end{equation}
where $\epsilon=(\ell/l_{k})^{d}$ and 
\begin{equation}
    \tilde{b}_{m}=\epsilon(1-\epsilon)^m \sum_{n=0}^{L_{\rm max}}\binom{m}{n}\frac{b_{n}\epsilon^n}{(1-\epsilon)^n}.
\end{equation}
Note that when $l_k=\ell$ Eq.~
\eqref{nonlocalop} reduce to~\eqref{opn}, and their local limit in the continuum
is also the standard d'Alembertian operator. But while the nonlocality of the continuum 
limit of~\eqref{opn} is restricted to scales of order $\ell$, that of~\eqref{nonlocalop} is of $O(l_k)$.

Again the operators $\tilde{B}^{(d)}$ can be represented by non-singular triangular matrices and can be 
inverted to obtain generalised nonlocal retarded Green functions parametrised by the nonlocality scale $l_k$.

\subsection{Sorkin Entropy}\label{sorkinent}

The final ingredient needed in order to explore the question of entanglement entropy 
of a quantum field on a causal set is a definition of entropy that does not require the notion of 
state at a moment of time.\footnote{Recall that we don't have access to the vacuum state 
$|0\rangle$ directly, but only via the two point function $\langle0| \phi\phi|0\rangle$~\cite{Sorkin:2017fcp}.}
A covariant definition of entropy for a Gaussian field,  
given purely in terms of spacetime correlators, was given in \cite{Sorkin:2012sn} and goes as follows.

Let $W$ and $i\Delta$ denote the Wightman and the Pauli--Jordan two point functions 
of a free (Gaussian) scalar field theory. Then, 
given a spacetime region $U$, the Sorkin entropy of the field in $U$ relative to its 
complement is given by
\begin{equation}\label{SEE}
\mathcal{S}(U)=\sum_\lambda \lambda \ln|\lambda|,
\end{equation}
where $\lambda$ are the eigenvalues of the generalised eigenvalue problem
\begin{equation}\label{geneigpr}
W|_{U}\, v=i\lambda\,\Delta|_{U}\,v\;\;\;\;\text{s.t.}\;\;\;\;R|_U\,v\ne0,
\end{equation}
The subscript $U$ indicates that the eigenvalue problem must be solved for the two-point functions 
restricted to pairs of points $(x,y)\in U$. Each eigenvalue $\lambda$ in~\eqref{SEE}  must be given its
correct multiplicity, which can be done by letting $v\sim u$ if $u = v + w$, where $w\in \kerr|_U$.

The fact that we have the condition $R|_Uv\ne0$, instead of the arguably more natural $\Delta|_Uv\ne0$, 
is because positivity of the Wightman function ($W\ge0$) only implies that 
$\kerr\subseteq \kerd$, the converse inclusion not being guaranteed in general.
So when $\kerr$ is a proper subset of $\kerd$ 
there exist $v\in\kerd$ for which $\lambda = \infty$ is the only solution to~\eqref{geneigpr}. 
When this happens we find an infinite entropy. 
This result can be traced back to normally distributed, purely classical components of the operator 
albegra $\A_U$. 
Strictly speaking therefore, when $\kerr\ne\kerd$ the entropy is not
an entanglement entropy. 
These last points can also be understood from the algebraic viewpoint of the QFT. 

To this end let $i=1,2,\dots,N$ be a natural labelling of the causal set $\CS$. We can think
of the field as a vector in $\mathbb{R}^N\ni\phi$, where each component 
$\phi^i$ is the value of the field at $i$.
Let $\A$ be the algebra generated by 
$\{\phi^i\,|\, i\in \CS\}$ acting irreducibly on a Hilbert space $\mathcal{H}$, together
with a global state $\rho$ such that $\langle A\rangle = \text{Tr}(\rho A)$ for all $A\in \A$.
Let $\A_U$ be the subalgebra generated by the set of operators 
$\Phi_U:=\{\phi^i\,|\, i\in U\}$, where $U\subset\CS$. 
$U$ will typically be chosen to be a causally convex  subset of $\CS$.
The commutant of $\Phi_U$, denoted $\Phi_U'$, is defined as the set of all operators 
that commute with all operators in $\Phi_U$.  From von-Neumann's bicommutant theorem
we have $\A_U = \Phi_U''$ and $\A_U' = \Phi_U'''=\Phi_U'$~\cite{Haag1996}. 

The centre of $\A_U$ is given by
$\mathcal{Z}_U=\A_U\cap \A_U'$, and consists of all operators that commute with $\A_U$.
An algebra is called a {\it factor} if its centre consists of only multiples of the 
identity $\mathcal{Z}_U=\mathbf{1}$. If $\A_U$ is a factor then it can be expressed as a factor 
of a tensor product, and the Hilbert space can be factorised into 
$\mathcal{H}=\mathcal{H}_U\otimes\mathcal{H}_{U'}$.
Note that in our case the $\A_U'$ is generated by $\{\phi^i\,|\, i\in \bar{U}\}$, where $\bar{U}$ is the  
spacelike complement of $U$. Triviality of the centre corresponds to the condition that $\kerr=\kerd$ in 
our language. This is because nontrivial operators in the centre are given by $v\cdot\phi$, for 
$v\in \kerd$, $v\notin\kerr$, which can be shown to be identically zero when $\kerr=\kerd$. Note also 
that when $\kerr\ne\kerd$ then no irreducible representation of $\A_U$ exists.
As we will see shortly this effectively leads to superselection sectors in $\mathcal{H}$.

In cases where the centre is not trivial one can still define a notion of entropy that 
reduces to standard entanglement entropy when the centre is trivial~\cite{PhysRevD.89.085012}.
The crucial step in the definition is to find a basis that diagonalises the centre. In this basis 
a generic element of the centre 
is given by 
\be
\begin{pmatrix} 
\lambda_1{\bf 1}_{d_1}& 0 & \dots & 0 \\
0 & \lambda_2{\bf 1}_{d_m} &\dots & 0  \\
\vdots & \vdots & & \vdots\\
0 & 0 & \dots  &  \lambda_m{\bf 1}_{d_m}
\end{pmatrix}
\ee
where each ${\bf 1}_{d_k}$ is a $d_k\times d_k$ identity matrix. In this basis the algebra 
$\A_U\cup\A_U' $ is isomorphic to  
\be
\begin{pmatrix} 
\A^1_U\otimes\A'^1_{U}& 0 & \dots & 0 \\
0 & \A^2_U\otimes\A'^2_U &\dots & 0  \\
\vdots & \vdots & & \vdots\\
0 & 0 & \dots  &  \A^m_U\otimes\A'^m_U 
\end{pmatrix},
\ee
and $\A_U$ is isomorphic to the block-diagonal representation of the full matrix algebra
\be
\begin{pmatrix} 
\A^1_U& 0 & \dots & 0 \\
0 & \A^2_U &\dots & 0  \\
\vdots & \vdots & & \vdots\\
0 & 0 & \dots  &  \A^m_U 
\end{pmatrix}.
\ee
The Hilbert space can be written as 
$\mathcal{H} = \oplus_k \mathcal{H}^k_{U}\otimes\mathcal{H}^k_{\bar{U}}$, each $k$
corresponding to a distinct superselection sector. 
We can perform a partial trace of the state $\rho_{U\bar{U}}$ over $\bar{U}$
\be
\rho_{U}:=\text{Tr}_{\bar{U}}(\rho_{U\bar{U}})=
\begin{pmatrix} 
p_1\rho^1_U& 0 & \dots & 0 \\
0 & p_2\rho^2_U &\dots & 0  \\
\vdots & \vdots & & \vdots\\
0 & 0 & \dots  &  p_m\rho^m_U
\end{pmatrix},
\ee
where $\text{Tr}(\rho^k_U)=1$ for all $k$, and $\sum_k p_k =1$.
The entropy of this state is then
\be\label{genee}
S(U) = -\text{Tr}	(\rho_U\log\,\rho_U) = -\sum_k p_k\log\,p_k - \sum_k p_k\text{Tr}(\rho^k_U\log\,\rho^k_U),
\ee
where the first term is the classical Shannon entropy and the second is a sum of the 
entanglement entropy for each sector weighted by the $p_k$. 
When elements of the centre are normally distributed random variables (as will be 
the case for our vacuum state) then the Shannon entropy term is known to be infinite,
unless some cutoff is introduced. 

To summarise, the Sorkin entropy of a spacetime region $U$ corresponds to the entanglement entropy 
of that region when $\kerr=\kerd$. Furthermore, it is equal to the entanglement entropy of $\Sigma$
when $\Sigma$ is a Cauchy surface of $U$.
If $\kerr\ne\kerd$ then the Sorkin entropy of $U$ 
is not a measure of entanglement entropy alone, but includes contributions coming from purely classical 
components of the Wightman function. These can also be understood from the algebraic viewpoint, 
where $\kerr\ne\kerd$ implies that $\A_U$ has a non-trivial centre, so that the Hilbert space cannot 
be split into a tensor product.\footnote{The
Hilbert space in fact becomes a direct integral of Hilbert spaces.} 
The centre's contribution to the entropy can be understood as the Shannon entropy of a classical random variable. This is infinite in the case of a gaussian theory.

\section{Sorkin Entropy with a Double Cutoff}\label{secsee}
\label{sec:SE}

Let us now consider a free, real scalar field living on a fixed causal set that is well-approximated by 
Minkowski spacetime of $d=2,3$ or 4 dimensions. The vacuum state of the field will be given implicitly in terms
of the Wightman two-point function, $W$, and is defined using the SJ-prescription described in Section~\ref{sec:EE-causet}.
We consider two classes of theories:
\begin{enumerate}
\item QFTs for which the Green functions are given by equations~\eqref{eq:loc}, which we shall refer to as ``local",
\item QFTs for which the Green functions are given by the inverse of nonlocal retarded d'Alembertian operators
\eqref{opn}, which we shall refer to as ``nonlocal".
\end{enumerate}
Following the the prescription of~\cite{Sorkin:2016pbz}, we begin by computing the entropy in the 
presence of two cutoffs on $W$. In later sections we will explore the physical significance
of these cutoffs.

\subsection{The setup}\label{secset}

\begin{figure}[h!]
\centering
\includegraphics[scale=0.2]{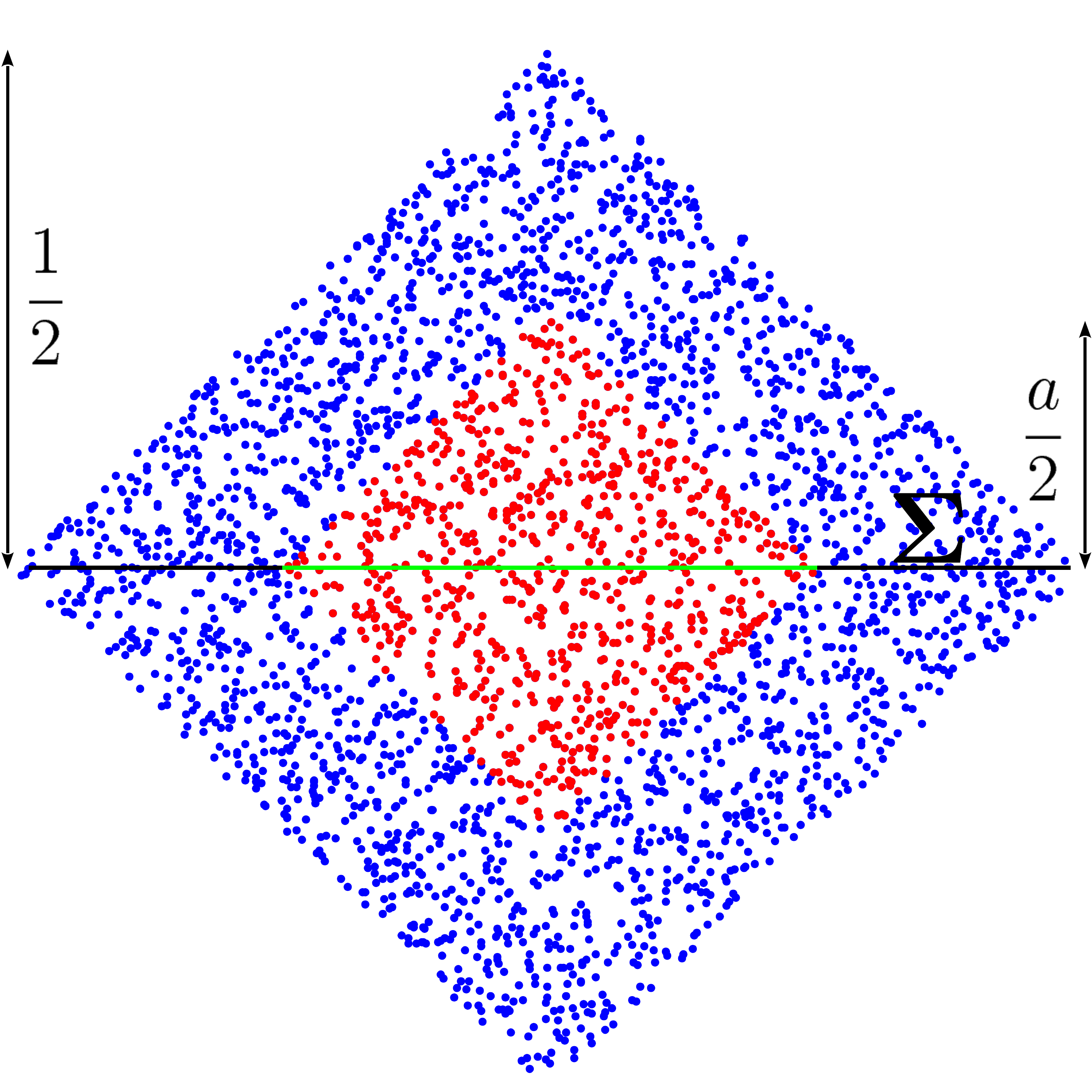}
\caption{Inner and outer diamonds for a sprinkling in 2$d$ Minkowski, with $N=2048$ and $V/V_U=4$. 
The Sorkin entropy $S(\CS_U)$ is the causet analogue of the entanglement
entropy of the vacuum state on $\Sigma$ restricted to the interior of the diamond.}\label{sprdiam}
\end{figure}

In Minkowski spacetime of dimension 2,3 and 4 we sprinkle into a causal diamond, $O$, centred
at the origin of the coordinate system and whose future and past most tips lie at 
$(1/2,{\bf 0})$ and $(-1/2,{\bf 0})$ respectively in all spacetime dimensions, see Figure \ref{sprdiam}. We define the SJ-vacuum of 
the sprinkled causet, $\CS_O$, using 
the prescription given in Section  \ref{sec:EE-causet} for both classes of Green functions. 
Within $\CS_O$ we select a subcauset, $\CS_U$, lying within a smaller 
diamond, $U$, also centred at the origin and whose future and past most tips lie at 
$(a/2,{\bf 0})$ and $(-a/2,{\bf 0})$ for some $0<a<1$ such that the ratio between the coordinate volumes of the outer and inner diamonds is fixed to be $V/V_U=1/4$.

Before computing the Sorkin entropy of $W_U$ we perform the double cutoff prescription described 
in~\cite{Sorkin:2016pbz}. The first cutoff is done on the global Wightman function, and is in effect a 
redefinition of $W$ following a truncation of $\Delta$. In particular, we define a truncated  
Pauli-Jordan function, $\Delta_\kappa$, by taking $\Delta$ and setting $\alpha \rightarrow 0$ 
whenever $|\alpha|\le\kappa$, $\alpha\in \text{spec}(\idelta)$ for some positive $\kappa$
(here $\text{spec}(\idelta)$ denotes the spectrum of $\idelta$).
We then define a regularised vacuum two-point function by $W_\kappa := \text{Pos(}\Delta_\kappa)$. 
The second cutoff is similar to the first, except that it is imposed inside the region where
the entropy is computed (in our case region $U$), and is performed on both $\Delta_\kappa|_U$ and
$W_\kappa|_U$ (equivalently $R_\kappa|_U$) simultaneously. As we will see later on, this cutoff is similar in spirit
to the cutoff one imposes in the continuum to make the entropy finite, 
since by choosing it appropriately one is in some sense excluding 
``transplanckian" modes from the computation. We will discuss their significance further in Section~\ref{cutoffs}.

Having made this double truncation we compute the Sorkin entropy of $U$
by first numerically solving the generalised eigenvalue problem~\eqref{geneigpr}
with $W_\kappa|_U$ and $i\Delta_\kappa|_U$, and then calculating
$S(U)=\sum_\lambda \lambda\log|\lambda|$.

\subsection{Selecting a Cutoff}

To be able to determine how the entropy scales as one varies the discreteness scale $\ell$, 
we must fix a cutoff on $\idelta$'s spectrum that has the appropriate scaling
with $N$. Consider for example the spectrum of $i\Delta$ in the continuum in a 2$d$ causal diamond of 
side length
$L$ given by $\alpha^{cont}=L/(2k)$, where for large integer $n$, 
the wavenumber, $k$, associated to each eigenmode of $\idelta$ is $k\sim\pi n/L$~\cite{Afshordi:2012aa} . The cutoff, both in the large diamond and the smaller subdiamond, 
is then implemented by setting to zero all eigenvalues of $i\Delta$ 
smaller than some minimum value, $\alpha^{cont}_{min}$, usually taken to correspond to the
minimal wavelength mode with 
maximum wavenumber $k_{\rm max}=2\pi/l_{\text{min}}$~\cite{Saravani:2013nwa}. 
One can think of the cutoff as effectively excluding modes whose wavelength is smaller
than some minimum wavelength from the calculation of the entropy. Note that 
even though this interpretation of the cutoff is not Lorentz invariant,
the procedure by which it is implemented is.

In order to translate this minimum eigenvalue in the continuum to the cutoff in the causal set, 
we simply identify the cutoff scale $l_{\textrm{min}}$ 
with the fundamental discreteness scale $\ell=\rho^{-1/d}$, and multiply by a factor of $1/\rho$
to get the dimensions right.\footnote{The origin of the mismatch between the dimensions of $i\Delta$ 
in the discrete and the continuum is due to the fact that, in the continuum theory $i\Delta$ is a 
Hermitian integral operator with mass-dimension $-2$ in every spacetime dimension (see~\cite{Saravani:2013nwa}). 
Whereas, in the discrete theory $i\Delta$ coincides with its would-be integral kernel in the continuum, 
which is just the commutator of two field strenghts, and as such it has spacetime dimension $d-2$, 
where $d$ is the spacetime dimension. Thus, a conversion factor with dimension $d$ is needed in order 
to compare the spectrum in the discrete with the one in the continuum.} Thus
\begin{equation}\label{2dcutoff}
\alpha_{\rm min}=\rho\,\frac{\sqrt{V}}{2k_{\rm max}}=\rho\frac{\sqrt{V}}{4\pi}\ell=\fr{\sqrt{N}}{4\pi},
\end{equation}
where we used $V=L^2$ and $\rho=N/V$. 

\begin{figure}[h]
\hspace{-0.8cm}
\centering
\includegraphics[scale=0.4]{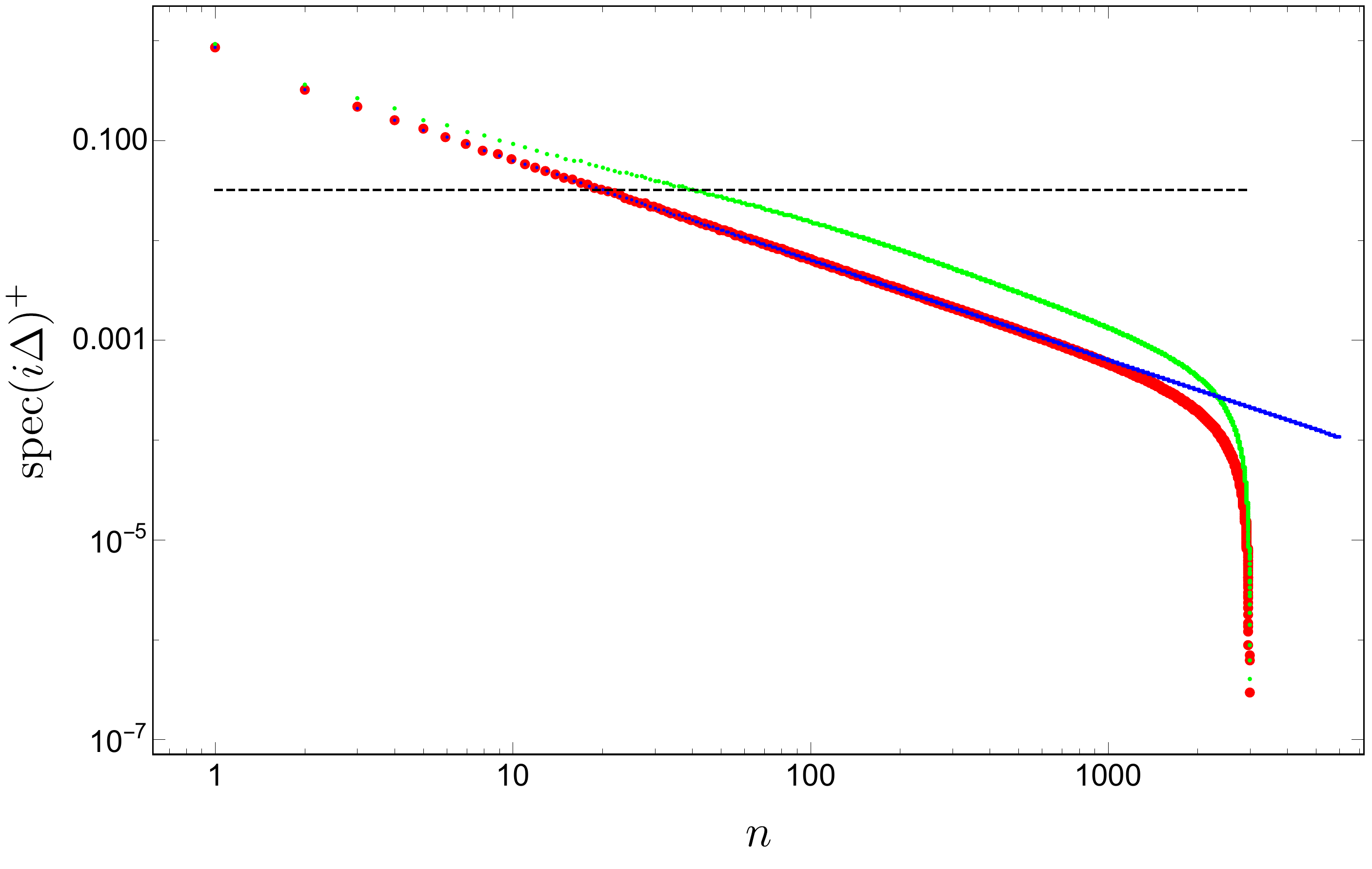}
\caption{Positive part of $\text{spec}(\idelta)$ for a diamond in Minkowski spacetime for both a sprinkling 
of the diamond (in red for the local theory and green for the nonlocal theory) and in the continuum (in blue). 
The eigenvalues have been ordered by size from largest to smallest. 
The red line represents the case discussed in the previous section of a local Pauli--Jordan function, 
the blue line is the spectrum of the continuum theory and the green one represents the spectrum of 
the nonlocal Pauli--Jordan function with $l_k=0.2$. The black, dashed line signals the end of the 
strictly infrared spectrum and the onset of the nonlocal regime.}\label{spectrum}
\end{figure}

Therefore, as we increase the sprinkling density $\rho$, the scale at which we implement the cutoff
increases as $\sqrt{N}$. This specific dependence of the minimal eigenvalue -- 
roughly corresponding to a mode of wavelength $\ell$ -- on $N$ is what ensures that as we vary $N$
we are consistently truncating the spectrum associated to modes of wavelength $<\ell$. 
Note that for a given $N$, this cutoff truncates the part of the spectrum that grossly deviates
from the continuum's spectrum, see Figure~\ref{spectrum}.
Furthermore, it is also worth noticing that the spectrum for the nonlocal theory, with $l_k\ne\ell$ in the UV, is offset by a factor of two
relative to the spectrum of the local theory. The difference between the two spectra begins
to emerge when eigenvalues are roughly of order $Ll_k$, where nonlocal effects become relevant. 
This implies that, even though the cutoff's dependence on $N$ is left unchanged (as we
will argue shortly), it differs from that of the local theory by an overall factor of 2
\begin{equation}\label{2dcutoff}
\alpha^{\rm (nl)}_{\rm min}=2\rho\,\frac{\sqrt{V}}{2k_{\rm max}}=\rho\,\frac{\sqrt{V}}{2\pi}\ell=\frac{\sqrt{N}}{2\pi}.
\end{equation}

A similar analysis in higher dimensions would require the knowledge of spectrum of $\idelta$ in the causal
diamond. However, since the cutoff is implemented in the UV, we are only
really interested in the scaling of eigenvalues of the UV modes with $k$. One can argue that these will also
go like $L/k$ (to leading order) in all dimensions, so that in $d$-dimensions we have 
\begin{equation}\label{Dcutoff}
\alpha_{\rm min}^{(d)}=c_d\,\rho\,\ell \,V^{\frac{1}{d}}=c_d\,\rho^{1-\frac{1}{d}}V^{\frac{1}{d}}=c_d\,V^{\frac{2}{d}-1}N^{1-\frac{1}{d}}.
\end{equation}
where coefficient $c_d$ parametrises our ignorance about the exact spectrum of $\idelta$ in the continuum, and 
in our simulations will be chosen so that it gets rid of the part of the spectrum that rapidly falls to zero, as in 2$d$.

Finally we must discuss the choice of cutoff for nonlocal theories. As we already noted for the 2$d$
nonlocal theory, the UV part of the spectrum differs from that of the local theory. 2$d$ is peculiar
in that the difference between the spectra of the local and non-local d'Alembertians, and therefore
of $\idelta$, in the UV in the continuum is merely a constant factor (see Equations~\eqref{UVexp}). 
In higher dimensions however, the functional dependence of the d'Alembertian on $k$ itself 
changes in the UV, and therefore so does that of the spectrum of $\idelta$. So the question 
becomes whether one should adapt the dependence of the 
cutoff on the power of $k$ (and therefore on $N$ via $k_{\rm max}=2\pi/\ell$) entering
the eigenvalues of $\idelta$ in the UV, or whether one should stick with the cutoff used for 
the local theories. 

The answer appears to be that if one wants to compare the two entropies
then the same cutoff should be chosen for both theories. Indeed, when one computes the entropy
for the nonlocal theories in the continuum using standard techniques (see Appendix~\ref{appA}), the same
cutoff is used (by construction) for both the local and nonlocal theories. It is clear that if one adapted the choice 
of cutoff to the d'Alembertian's dependence on $k$ in the UV then the scaling of the entropy would change,
and a direct comparison with the local result would be meaningless. Thus, in our simulations
we have used the same cutoff, given by Equation~\eqref{Dcutoff}, for both the local and nonlocal
theories in every dimension.

\subsection{Numerical Results}

\begin{figure}[h] 
     \includegraphics[width=1\linewidth]{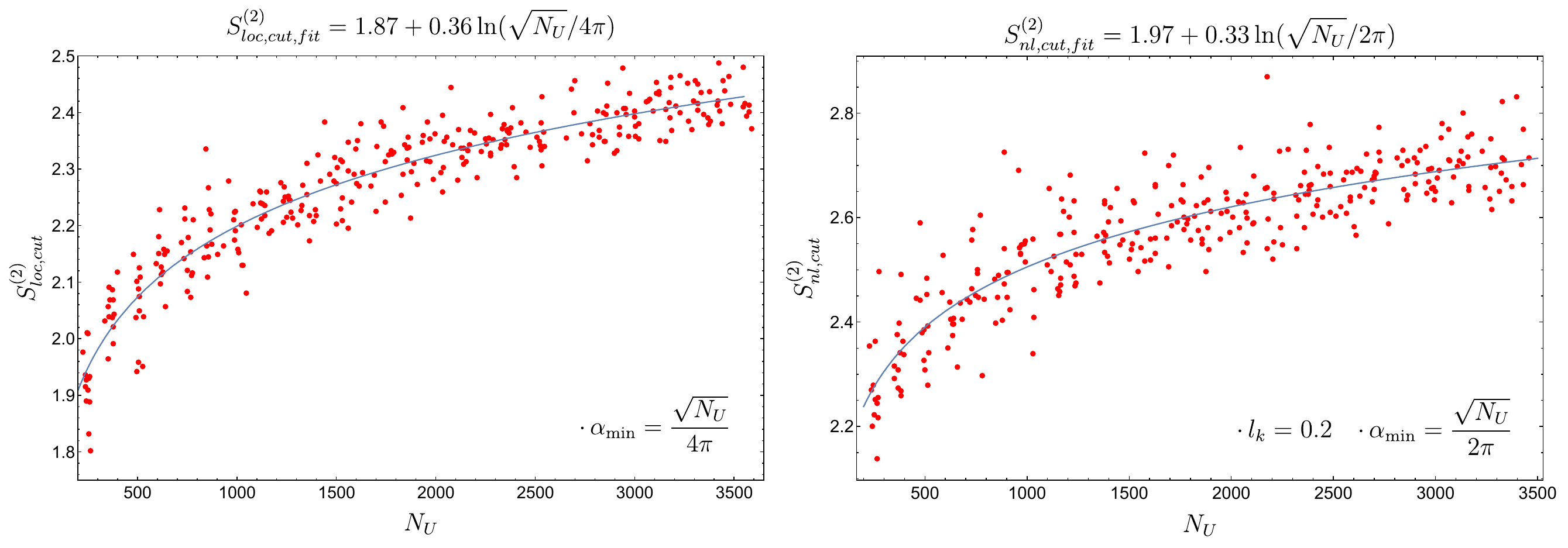} 
\caption{Entanglement entropy vs. number of points in the subregion $U$ for the local (left) and nonlocal (right)
theories with cutoffs on 
both the global Wightman function, $W$, and $W_\kappa|_U$.}   
\label{arealaw2d}   
\end{figure}

\begin{figure}[h] 
    \includegraphics[width=1\linewidth]{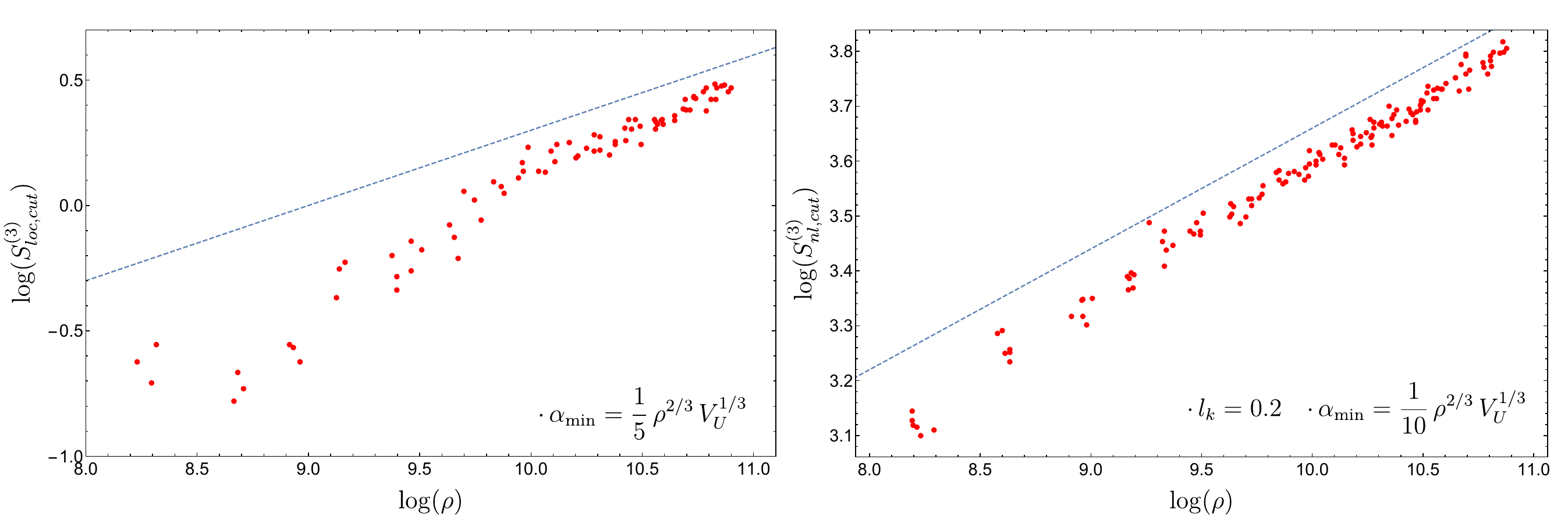} 
    \caption{Entanglement entropy vs. number of points in the subregion $U$ for the local (left) and nonlocal (right) theories with cutoffs on both the global Wightman function, $W$, and $W_\kappa|_U$. The blue dashed lines have slopes $\rho^{1/3}$ (left) and $\rho^{2/9}$ (right), they correspond to the ``area-law'' scaling expected in the local theory and in the nonlocal theory (as for Eq.~\eqref{ent_entropy}) respectively. }
     \label{arealaw3d}
\end{figure}

\begin{figure}[h] 
    \includegraphics[width=1\linewidth]{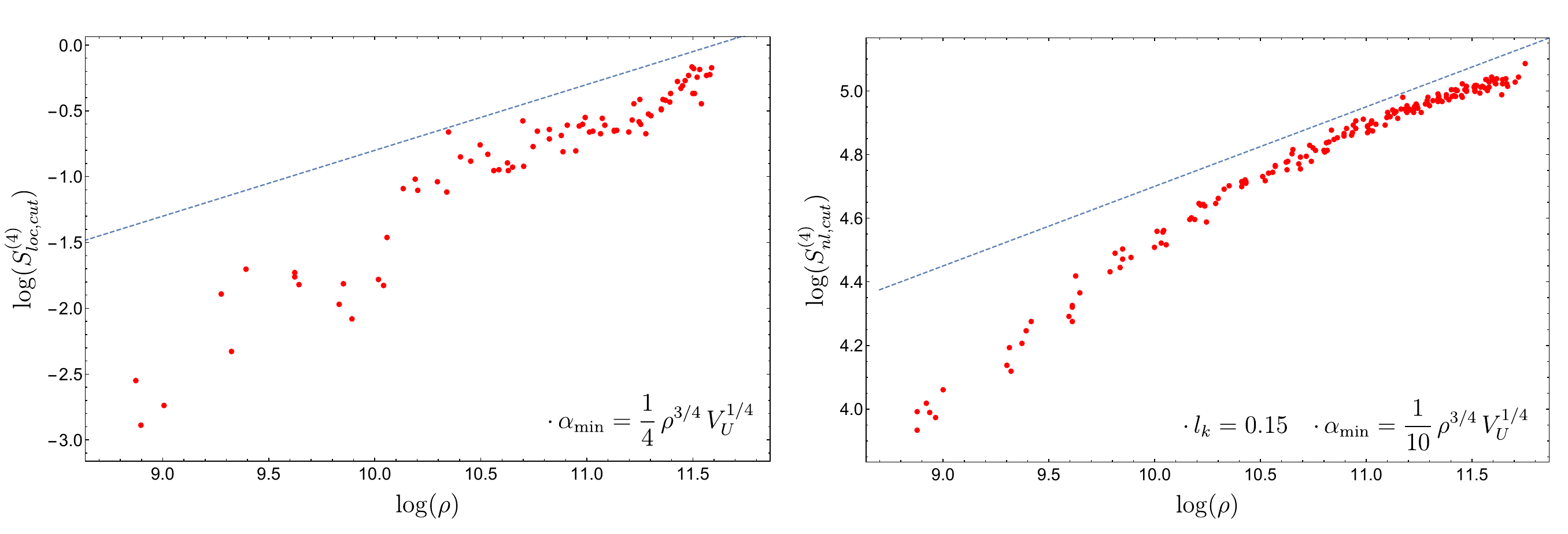} 
    \caption{Entanglement entropy vs. number of points in the subregion $U$ for the local (left) and nonlocal (right) theories with cutoffs on both the global Wightman function, $W$, and $W_\kappa|_U$. The blue dashed lines have slopes $\rho^{1/2}$ (left) and $\rho^{1/4}$ (right), they correspond to the ``area-law'' scaling expected in the local theory and nonlocal theory (as for Eq.~\eqref{ent_entropy}) respectively.}
     \label{arealaw4d}
\end{figure}

With functional forms for the cutoffs in all dimensions at hand, we can now proceed to compute the 
entanglement entropy of the region $U$. In each case we implement the cutoff given in~\eqref{Dcutoff}
in the large diamond, and the same cutoff, with $V$ replaced by $V_U$, in the small diamond. The parameter setting the size of the inner diamond is chosen to be $a^{-1}=4,2^{5/3},2\sqrt{2}$ for $d=2,3,4$ respectively.
Results are shown in figures~\ref{arealaw2d},~\ref{arealaw3d}
and~\ref{arealaw4d}. For each dimension we show plots
for both the local theory and the nonlocal theory with $l_k=0.2$ in 2 and 3 dimensions and $l_k=0.15$ in 4$d$.

Consider first the 2$d$ results shown in~\ref{arealaw2d}. Note how both the local and nonlocal theories
are in good agreement with the known continuum result~\cite{calabrese2009entanglement}, including the overall coefficient of $1/3$.

Next consider the 3$d$ and 4$d$ data shown in figures~\ref{arealaw3d} and~\ref{arealaw4d}. 
Here our data is not as conclusive as in 2 dimensions, something which can most likely
be traced back to the fact that the sprinkling densities in these dimensions are much smaller
than in 2$d$, with the 4$d$ attaining the smallest densities altogether. Nonetheless, there
are strong hints that the scaling at the highest densities is approaching an 
area law, at least for the local theory.
Indeed recall that in $d>2$ the area law is
\be
S(U)\sim \fr{L^{d-2}}{l^{d-2}}\sim L^{d-2}\rho^{1-2/d},
\ee
where $L$ is a length scale associated to the boundary of the subregion. 
One would therefore expect the slope in a log-log plot against $\rho$ to asymptote $1-2/d$ as $N$ gets large. 
While our numerical data for the local theory appears to be consistent with this limit, in the nonlocal theory 
the slope is less pronounced, suggesting that
the scaling of entropy with $\ell$ is weaker than an area law.
As we will see shortly, this turns out to be consistent with continuum calculations
for the entanglement entropy of nonlocal field theories parametrised by a scale $l_k$, 
when $l_k$ differs from  the cutoff scale $\ell$ needed to render the entanglement entropy finite.

Modulo this caveat for the nonlocal theories, that will be further discussed in the next section, one can
say with some degree of confidence that the numerical simulations for the Sorkin entropy on causal sets 
{\it with} two cutoffs, is consistent with the area law in the dimensions discussed. Obviously, a more definitive
statement on this would require 
simulations to higher densities, and we are hopeful
that we will be able to achieve this in the near future using the new causal sets generator introduced
in~\cite{cunningham2017causal}.

\subsection{Entanglement Entropy for Nonlocal QFTs in the Continuum}

The continuum limit of the Green functions used in the construction of our causet QFTs are known
to be Green functions of nonlocal, retarded d'Alembertian operators~\cite{Belenchia:2014fda}
(except for the 2$d$ local theory, whose continuum limit is the exact retarded Green function for $\Box$).
One could ask therefore how our causet results for the entanglement entropy compare with
the entanglement entropy of their respective nonlocal field theories in the continuum. 

To that end note that the entanglement entropy of (free) nonlocal field theories in the continuum in 
$d$-dimensions is given by~\cite{Nesterov:2010yi}
(see Appendix~\ref{appA} for further details)
\begin{equation}\label{ent_entropy}
S=\frac{A(\Sigma)}{12 (4\pi)^{(d-2)/2}}\int_{\epsilon^2}^\infty \frac{ds}{s}\; P_{d-2}(s),
\end{equation}
where $\epsilon$ is a UV cutoff that makes the integral finite ($s$ has dimensions of a length squared) and
\begin{equation}\label{coneheatk}
P_{d-2}(s)=\frac{2}{\Gamma(\frac{d-2}{2})}\int_0^\infty dp \;p^{d-3}\;e^{-sF(p^2)},
\end{equation}
where $F(p^2)$ is the Fourier transform of the d'Alembertian operator. 
For the class of nonlocal field theories parametrised by $l_k > \ell$, $F(p^2)$ depends on $l_k$, so that
an $l_k$ dependence will in general enter the area law~\eqref{ent_entropy}.  We will also
set $\epsilon = \ell$ since we assume that the fundamental cutoff is given by the 
fundamental discreteness set by the causal set.

Even though a closed form solution of~\eqref{coneheatk}, and therefore~\eqref{ent_entropy}, does not exist, 
one can find the leading order UV contribution to the entropy analytically. 
By using the UV limit of $F$ in $d=2,3$ and 4 dimensions, we find that in the limit $p^2\gg l_k^{2}$, 
\begin{equation}\label{nlent}
\begin{split}
&S^{(2)}_{UV}\propto \ln \left(\fr{L}{\ell}\right),\\
&S^{(3)}_{UV}\propto\frac{A(\Sigma)}{\ell^{2/3}\,l_k^{1/3}},\\
&S^{(4)}_{UV}\propto\frac{A(\Sigma)}{\ell\,l_k}.
\end{split}
\end{equation}
(The interested reader is invited to look at Appendix~\ref{appA} for more details).
A few comments are in order.

First note that in $d=3,4$, the scaling of the entropy with respect to the cutoff $\ell$ is weaker than in the 
local case due to the presence of the nonlocality scale $l_k$ in the UV expansion of d'Alembertian (Eq.~\eqref{UVexp}). 
While in $d=2$, the nonlocality scale does not enter the UV expansion of the wave operator, 
hence the leading contribution to the entanglement entropy in the UV is unchanged relative to the local theory.

Secondly, even though the continuum limits of the 3 and 4$d$ local theories on the causal set are also nonlocal,
the divergence of the (UV limit of the) entanglement entropy in these theories is left unchanged  
with respect to the local theory, because the nonlocality scale in these cases {\it is} the cutoff scale $l_k=\ell$.
This fact is further confirmed by numerical simulations of the nonlocal theories with $l_k = l$, which show
a stronger divergence with respect to the case $l_k>\ell$.

This analysis confirms that for $l_k>\ell$ the entanglement entropy 
should diverge less rapidly than the usual area law as $\ell\rightarrow 0$, and appears to be confirmed by 
our numerical results which are consistent with the scalings~\eqref{nlent}, in the asymtptotic limit. 
Again, larger simulations are needed to to confirm the asymptotic regime has indeed been reached.

\section{Sorkin Entropy: Local and Global Cutoff}
\label{cutoffs}

Having discussed properties of the Sorkin entropy in the presence of two cutoffs, one might wonder 
why the need to introduce these cutoffs at all, given that the whole idea behind computing entanglement entropy
on causal sets was to get a finite entropy {\it without} the need to introduce a cutoff. 
This question is a valid one, and in the next two sections we will study the entropy on causal
sets without any cutoffs, as well as with the two cutoffs implemented individually, 
in order to try and get a better understanding
for why they are needed and what it is that they do. 
(In the sections that follow, whenever we refer to, or show results of, numerical simulations, 
they are for the local theory in 2$d$. But the general arguments also hold in other dimensions 
and for the class of nonlocal field theories.)

\subsection{The Need for a Local Cutoff}\label{loccut}
\begin{figure}[h!]
\centering
\includegraphics[scale=0.3]{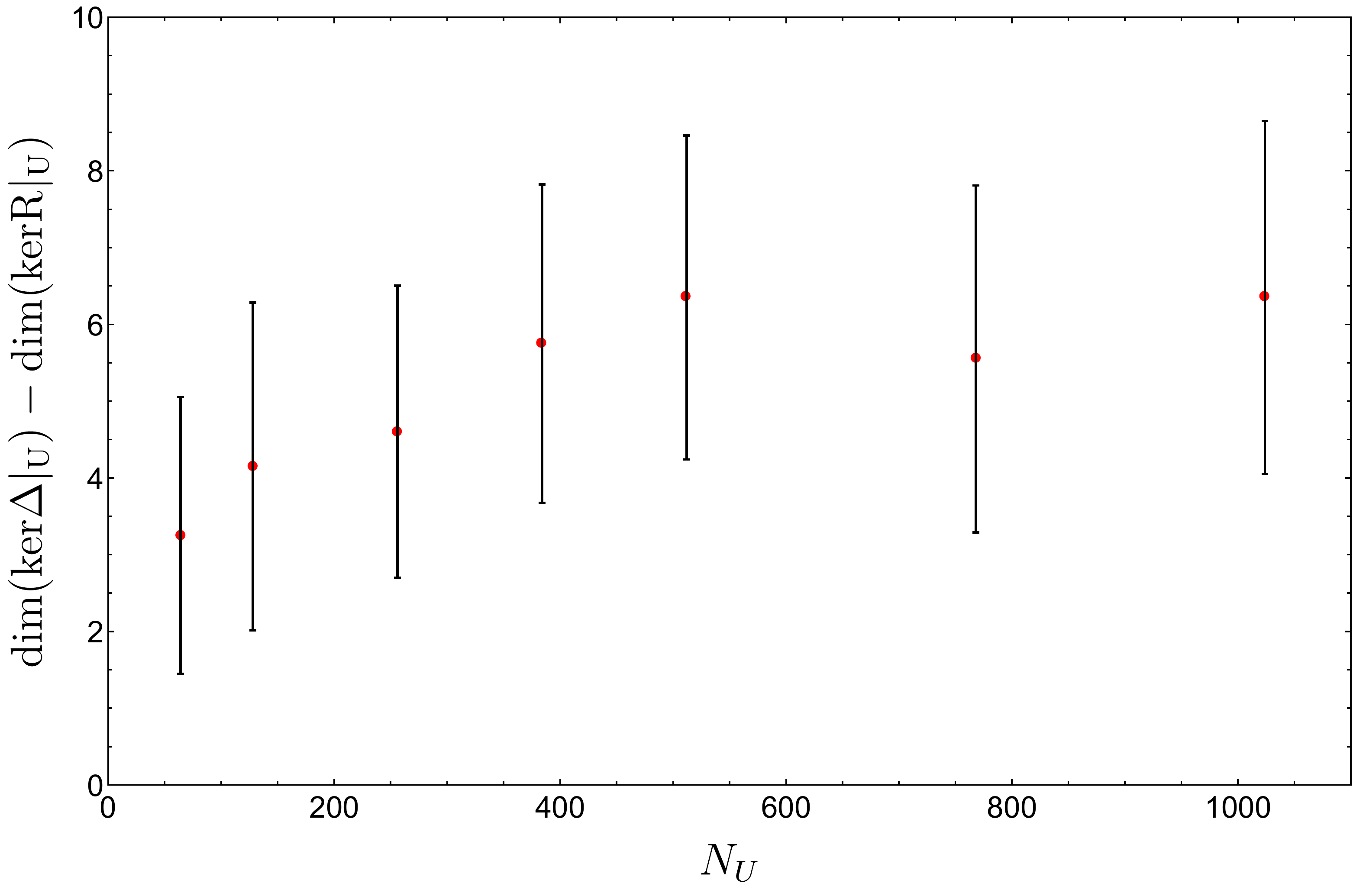}
\caption{Plot of the average difference between $\kerd|_U$ and $\kerr|_U$.}\label{RvsD}
\end{figure}

Consider again the set-up of Fig.\ref{sprdiam}.
We begin by computing the Sorkin entropy of $W_U$ {\em in the absence of any cutoff}.
The first thing to note is that in subregion $U$, $\kerr|_U \ne \kerd|_U$ in general, see Figure \ref{RvsD},
so that {\em the entropy is actually infinite}.
Now recall that these infinities are different from the one that appears in the continuum, 
in that they come from classical components of the Wightman function, and not from 
correlations over arbitrarily small scales. This immediately suggests that the entropy can
be made finite by implementing a cutoff on $W|_U$ in such a way as to augment the 
kernel of its real part $R|_U$. In practice this can be done by imposing that $v\in\kerr$ if $v\in \kerd$, which
leads to a new Wightman function in $U$ that we will denote by $\tilde{W}|_U$.

We can also understand the cutoff on $W$ as a deformation of the algebra $\A_U$. Before 
the cutoff the (pre-deformed) algebra $\A_U$ is fully characterised by
$\{\phi^i\,|\,i\in \CS_U\}$, the quadratic relations $\idelta^{ij}_U=[\phi^i,\phi^j]$ and the linear
relations $u\cdot\phi=0$, $u\in\kerr_U$.
But after the deformation the set of linear relations is enhanced to $v\in\kerd|_U$~\cite{Sorkin:2012sn}. 

\begin{figure}[h!]
\begin{minipage}{.5\linewidth}
\centering
\subfloat[]{\label{main:a}\includegraphics[width=1\linewidth]{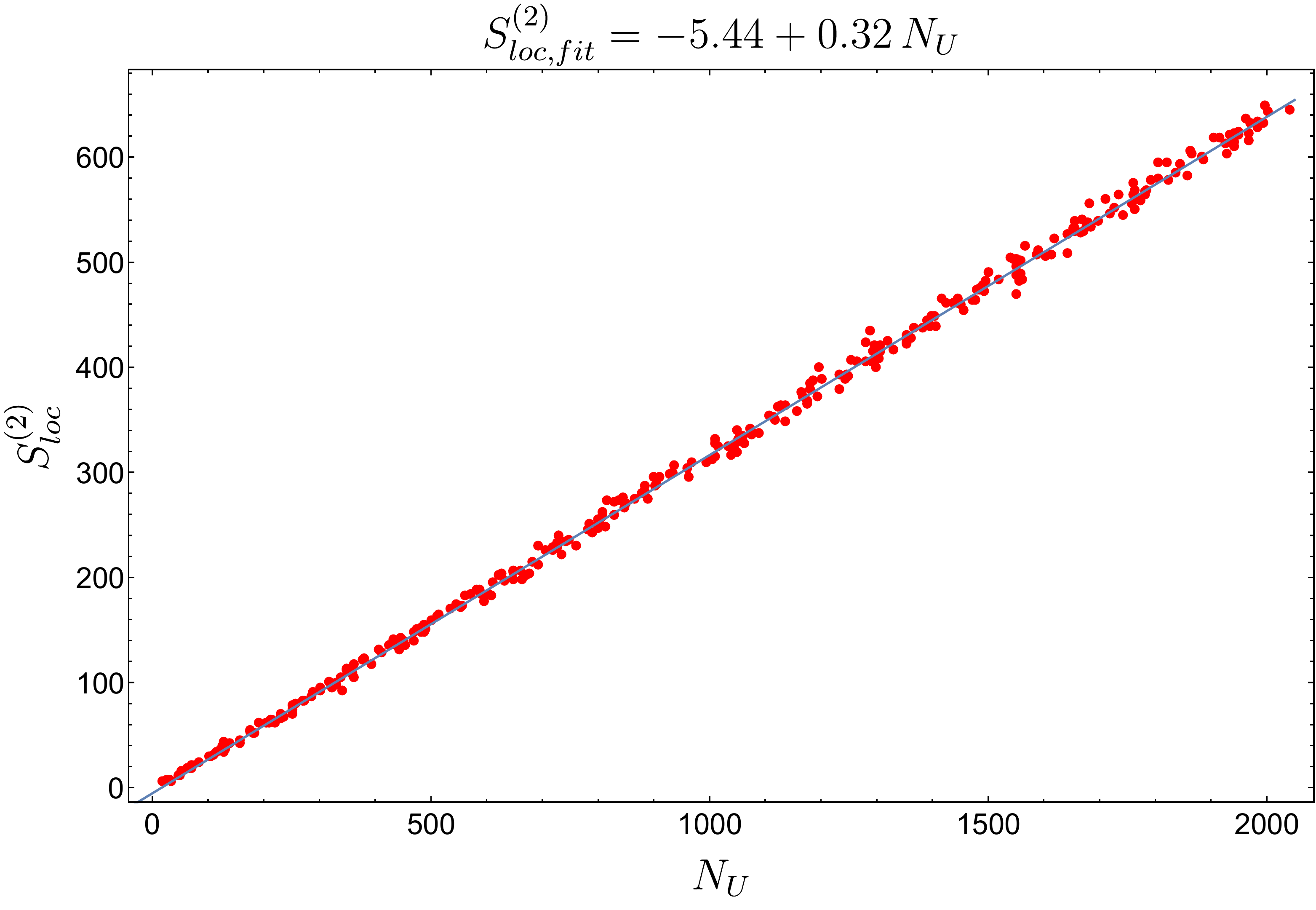}}
\end{minipage}%
\begin{minipage}{.5\linewidth}
\centering
\subfloat[]{\label{main:b}\includegraphics[width=1\linewidth]{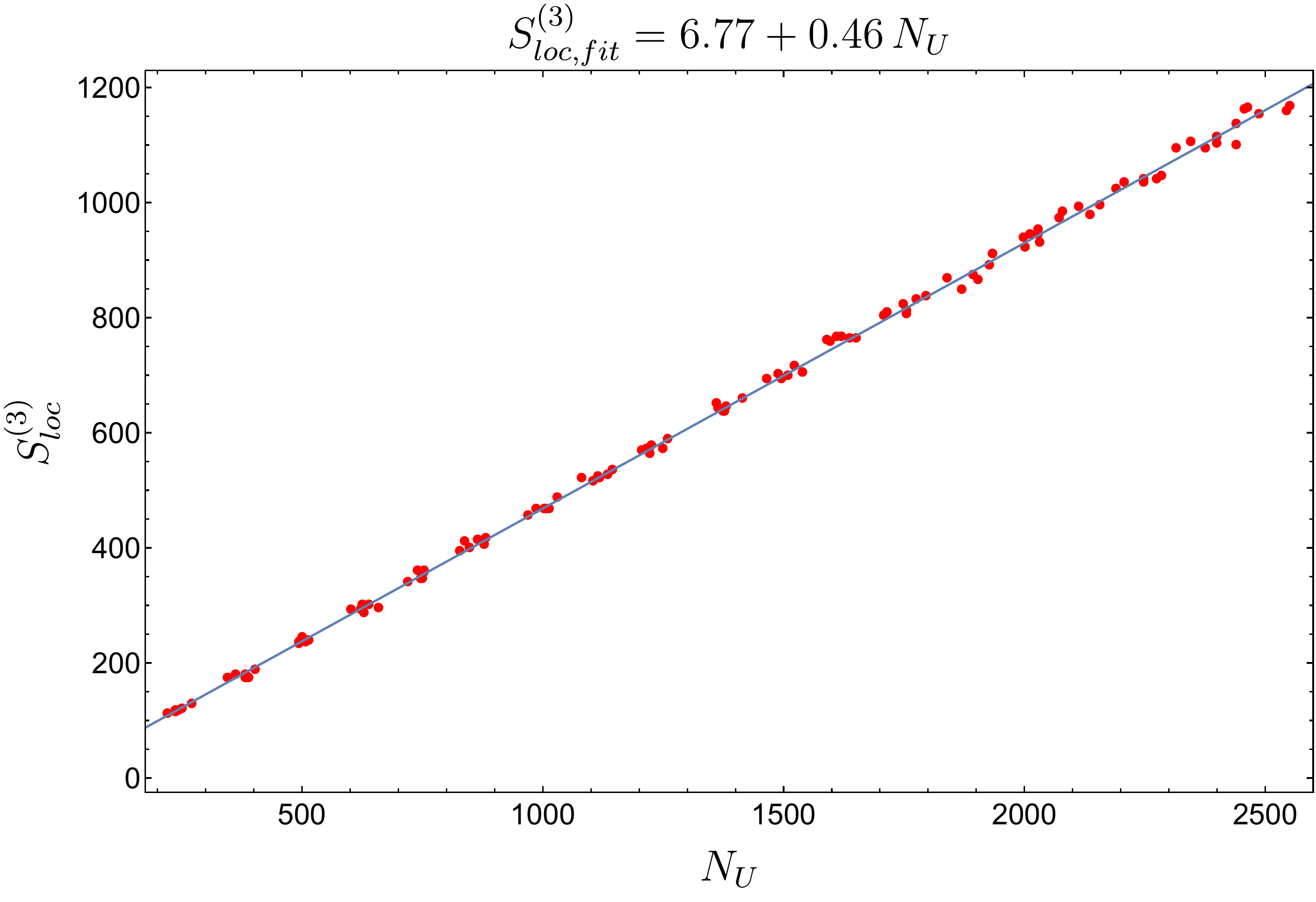}}
\end{minipage}\par\medskip
\begin{center}
\begin{minipage}{.5\linewidth}
\centering
\subfloat[]{\label{main:c}\includegraphics[width=1\linewidth]{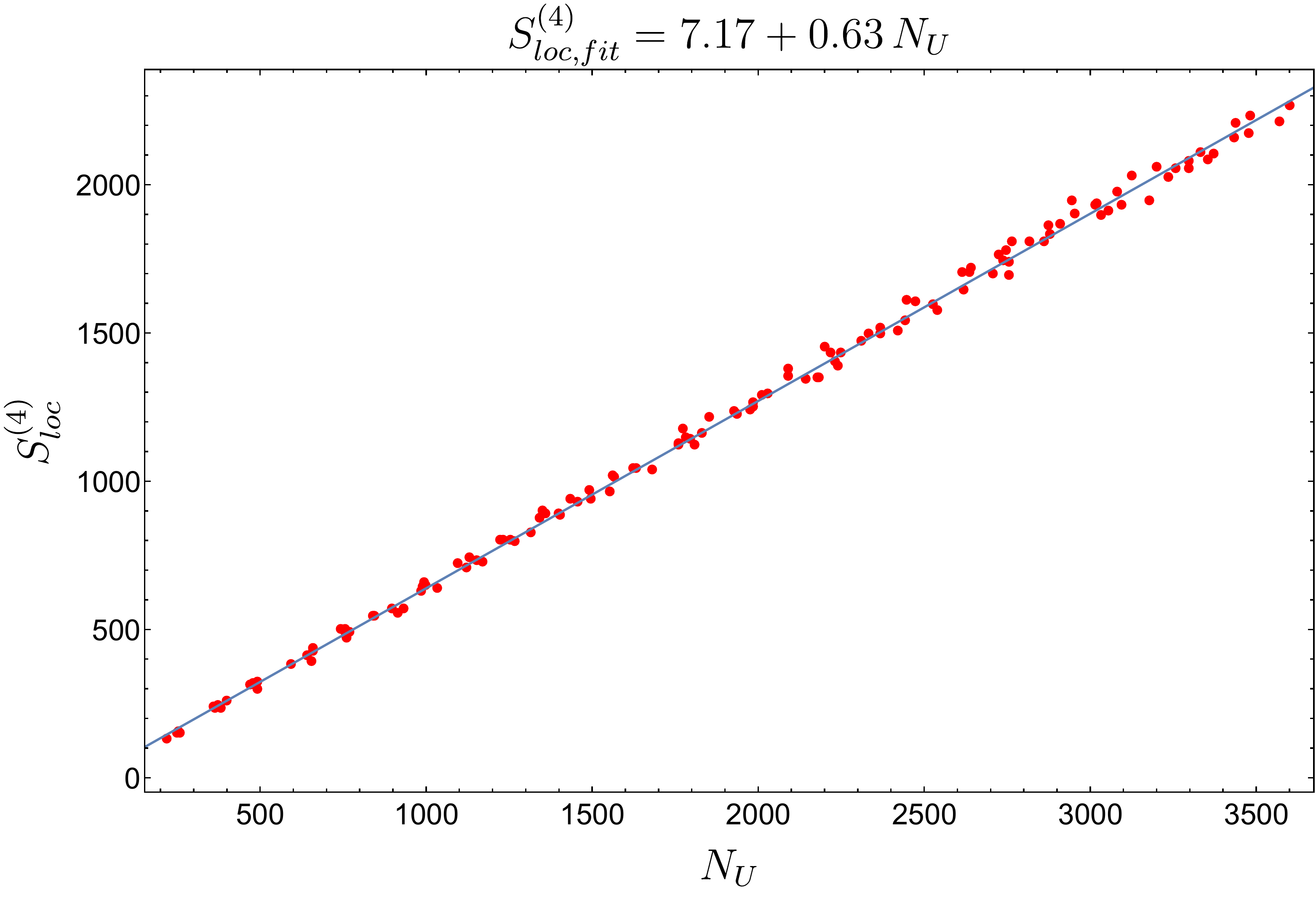}}
\end{minipage}
\end{center}
\caption{Finite part of the entanglement entropy vs. number of points in the subregion $U$ for d$=2$ (left), 3 (right) and 4 (bottom). This finite contribution is obtained by matching the kernel of $R$ to the one of $\Delta$ and follows a spacetime volume law.}
\label{locvol}
\end{figure}

\begin{figure}[h!]
\begin{minipage}{.5\linewidth}
\centering
\subfloat[]{\label{main:a}\includegraphics[width=1\linewidth]{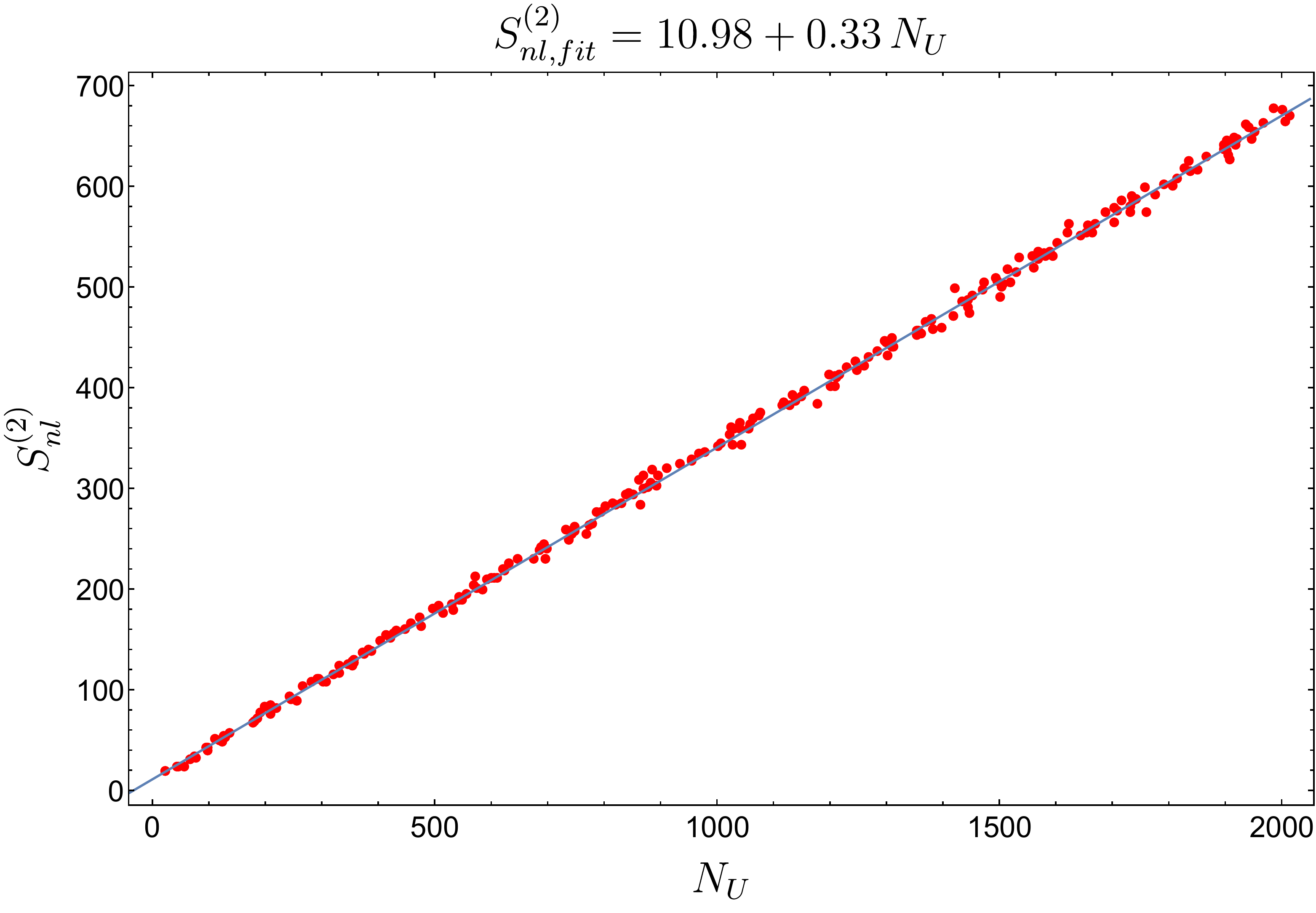}}
\end{minipage}%
\begin{minipage}{.5\linewidth}
\centering
\subfloat[]{\label{main:b}\includegraphics[width=1\linewidth]{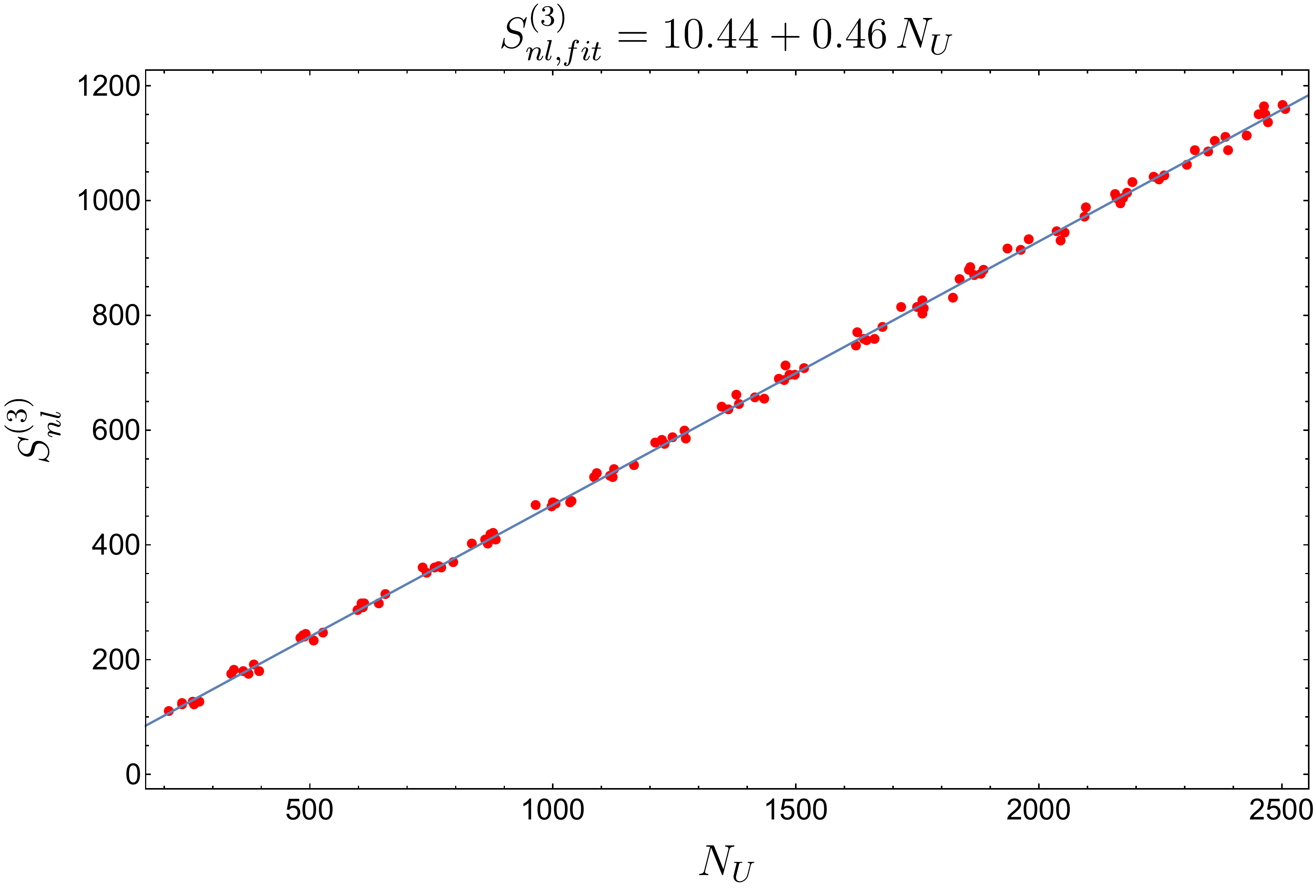}}
\end{minipage}\par\medskip
\begin{center}
\begin{minipage}{.5\linewidth}
\centering
\subfloat[]{\label{main:c}\includegraphics[width=1\linewidth]{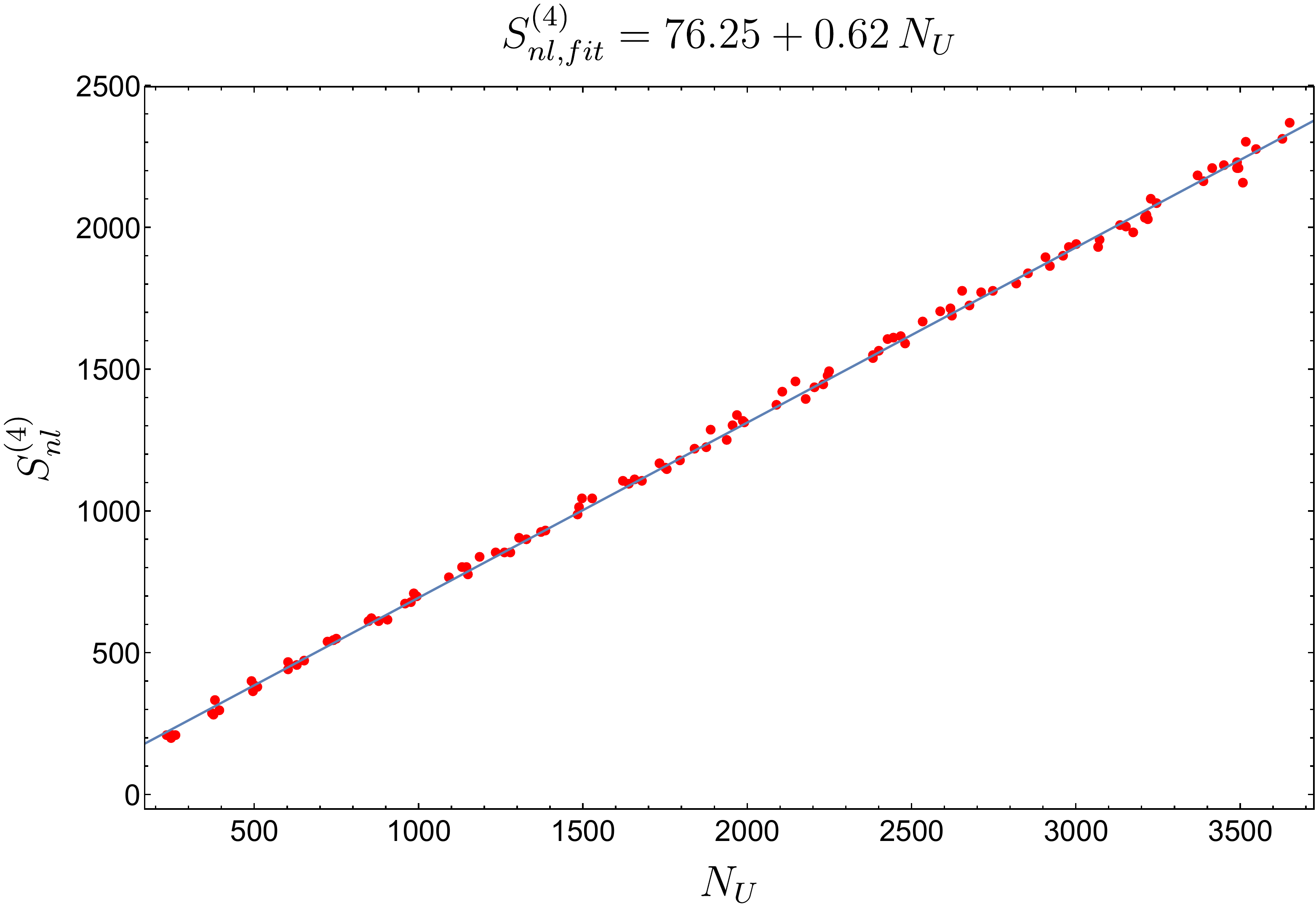}}
\end{minipage}
\end{center}
\caption{Finite part of the entanglement entropy vs. number of points in the subregion $U$ for d$=2$ (left), 3 (right) and 4 (bottom), in the nonlocal case. This finite contribution is obtained by matching the kernel of $R$ to the one of $\Delta$ and follows a spacetime volume law.}
\label{nlvol}
\end{figure}

Having defined $\tilde{W}|_U$ we can recalculate the Sorkin entropy of $U$.
Figures~\ref{locvol} and~\ref{nlvol} show plots of the Sorkin entropy in 2,3 and 4 dimensions for the local
and nonlocal theories respectively. 
In all cases we find that the entropy scales like the spacetime volume of the region, which is consistent
with the results found in~\cite{Sorkin:2016pbz}.
\footnote{Note that  even though the entropy of $U$ is now finite, the deformed algebra $\A_U^{cut}$ still has a non-trivial 
centre, since $\kerr|_{\bar{U}}\ne \kerd|_{\bar{U}}$ where $\bar{U}$ is the spacelike complement of $U$.
Therefore one would have to impose a cutoff also on $W_{U'}$ in order to ensure that both the entropies
of $U$ and $\bar{U}$ are finite and that the Hilbert space can be written as a tensor product 
$\mathcal{H}_U\otimes\mathcal{H}_{\bar{U}}$. However, since neither $W|_U^{cut}$ nor
$W|_{\bar{U}}^{cut}$ correspond to restrictions of $W$ to their respective regions or, equivalently,
$\A_U^{cut}$ and $\A'^{cut}_{U}$ are not subalgebras of $\A$,
the state whose partial traces we are computing the entropy of, i.e. 
$\rho_{U\bar{U}}\in\mathcal{H}_U\otimes\mathcal{H}_{\bar{U}}$,
will not in general satisfy $\text{Tr}(\rho_{U\bar{U}} A)=\text{Tr}(\rho A)$ for 
$A\in \A_U\cup \A'_{U}$. In fact $\rho_{U\bar{U}}$ 
is not even a state in the algebra $\A_U\cup \A'_{U}$, but rather is a state 
in $\A^{cut}_U\cup \A'^{cut}_{U}$). We will come back to the question of 
what the global state for which the restriction to $U$ and $\bar{U}$ gives this entropy is,
and how one should interpret it, in Section \ref{secdisc}. }

The cutoff used to define $\tilde{W}|_U$ made the entropy finite 
by deforming $W$ just enough so that $\kerr= \kerd$. 
But one can now ask what happens if the cutoff is increased, so as to augment 
both $\kerr$ and $\kerd$ further, while preserving that $\kerr= \kerd$. 
Looking back at the spectrum of $\idelta|_U$ (see e.g. Figure~\ref{spectrum}), 
one can see that the cutoff can be chosen so as to eliminate that part of  $\text{spec}(\idelta|_U)$
that deviates from the continuum's spectrum altogether. If we do this then the entropy of $U$ 
scales differently with $\ell$. In particular, in 2$d$ the scaling law
is actually logarithmic (but with a coefficient that is greater than $1/3$), while in higher
dimensions the scaling improves but not to the extent that one recovers an area law 
(at least not at the densities explored in our simulations).

We can think of the cutoff in $U$ as serving a double purpose: first and foremost it renders
the entropy finite; and secondly it can be chosen so as to eliminate part of the spectrum of $\idelta$ that
one would associate to transplanckian modes in the continuum.
As we will see in the next section this cutoff alone does not help when computing the entropy
of regions whose entropy is known to vanish in the continuum. This will indicate the need for a second
cutoff on the global Wightman function.

\subsection{The Need for a Global Cutoff}

Thus far in our analysis we have restricted our attention to the setup described in Section~\ref{secset}, 
but one can compute the Sorkin entropy for other subregions too. It turns out that, much
like for region $U$ discussed previously, $\kerr|_\cau \ne \kerd|_\cau$ is generally true for any proper 
subregion, $\cau$, of the causal set. Thus, the entropy of any subregion is infinite.
 
\begin{figure}[h!]
\centering
\includegraphics[scale=0.4]{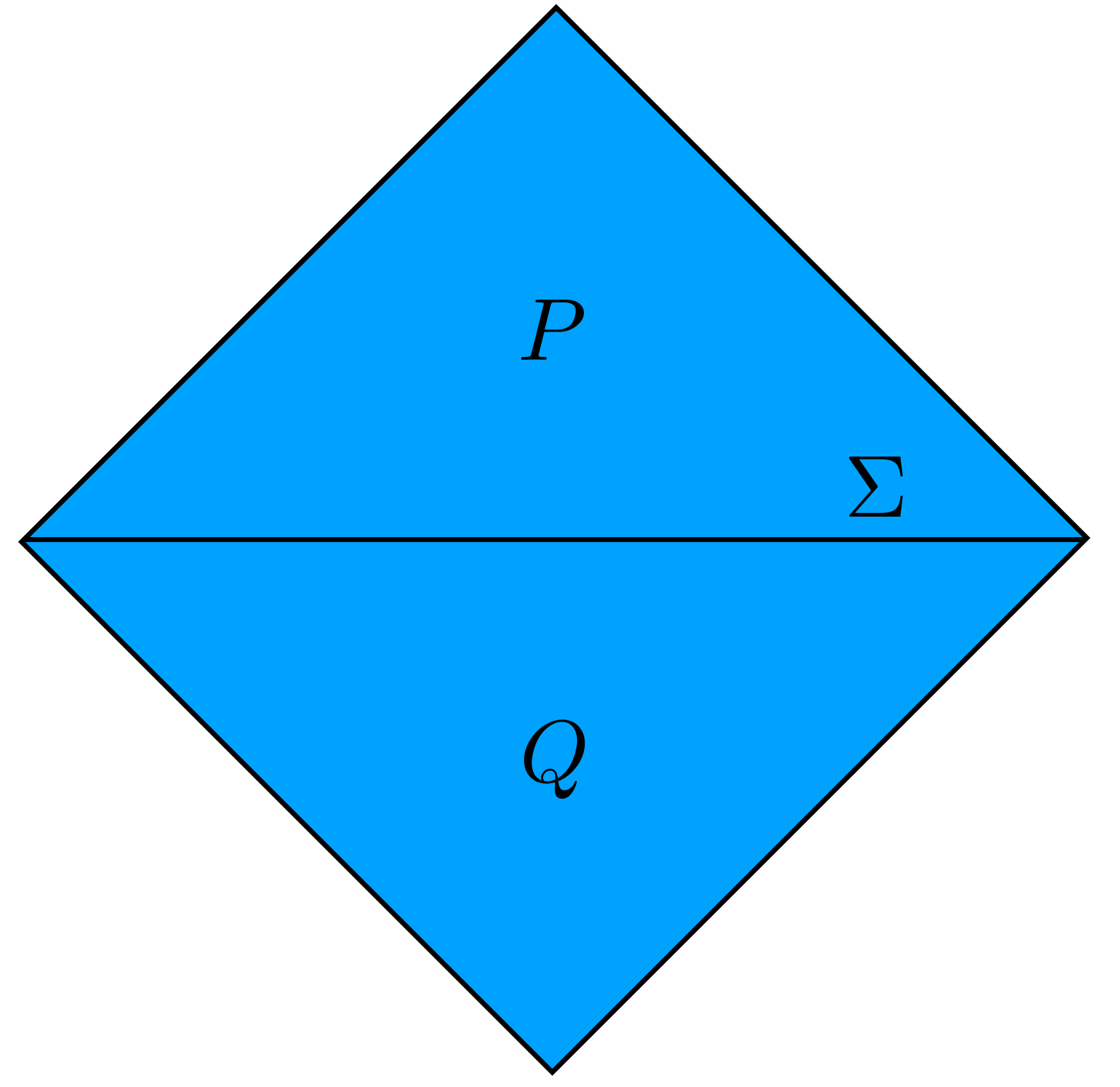}
\caption{Causal diamond in $\mathbb{M}^2$ with Cauchy surface $\Sigma$. $P$ and $Q$ are the future and past domains
of dependence of $\Sigma$ respectively.}\label{splike}
\end{figure}

A particularly interesting class of regions are what we refer to as ``Cauchy regions". In the continuum
these are simply subregions, $\cau$, such that $D^+(\cau)\cup D^-(\cau)$ is the full spacetime, and one can think
of them as thickenings of Cauchy surfaces. Cauchy regions in causal sets are similarly defined.
For example, Figure~\ref{splike} shows two particular Cauchy regions defined as the 
future, $D^+(\Sigma)$, and past, $D^-(\Sigma)$, Cauchy domains of dependence of the Cauchy surface 
$\Sigma$ defined by $t=0$.\footnote{In the causal set one would have to 
define this partition in terms of the causal set alone and not with reference
to a property of the continuum spacetime we sprinkled into, such as the surface $t=0$. This can be easily
done by using a maximal antichain as the analogue of a Cauchy surface, with regions $\CS_P$ and $\CS_Q$ then
given by the future and past of the maximal antichain respectively. 
For the purposes of our analysis this distinction will not be important,
but the reader should keep in mind that the setup can be made to be fully independent of the 
sprinkled region.}

In the continuum the entropy
of this region must vanish, since the algebra of $P:=D^+(\Sigma)$, $\A_P$, is equivalent to the full algebra $\A$
by virtue of the equations of motion.\footnote{In fact,
even the (formal) algebra generated by operators $\phi$ and $\dot{\phi}$
on $\Sigma$ is equal to $\A$, due to the second order, hyperbolic 
nature of the equations of motion.} In a sense this is what it means to be a Cauchy region, in the continuum.
But as we discussed already, in the causal set we are faced with the problem that in these subregions 
$\kerr|_{P,Q}\ne\kerd|_{P,Q}$ in general, so that their entropy is infinite.

However, introducing a cutoff like the one imposed in $U$ does not actually enforce the vanishing of this entropy. 
While it does render it finite, it still scales like $1/\ell^d$ and therefore diverges in the limit $\ell\rightarrow0$.
Furthermore, increasing the cutoff does not help, because it 
merely weakens the divergence of $S$ with $\ell$ without eliminating it altogether,
unless one is willing to throw away the whole spectrum of $\idelta$. 

This peculiar property of the entropy on the causal set appears to be symptomatic of 
poorly defined equations of motion.
Indeed, recall that when $\kerd=\kerr$, as is the case for the
global Wightman $W$, $\kerd$ {\it defines} the equations of motion via the linear relations
\be\label{eom}
A^T\phi = 0,
\ee
where $A$ is defined as
\be
A = (v_1\;v_2\;\dots v_d),
\label{eq:A}
\ee 
and $v_\alpha$, $\alpha=1,\dots,k=\text{dim}(\kerd)$ is a basis for $\kerd$.
But $k$ is typically very small, and grows very slowly with $N$; 
for example in 2$d$ numerical simulations suggest that 
$k\sim O(\log\,N)$ (c.f. Figure~\ref{RvsD}). This implies that only regions whose
size is $\,\gtrsim N-\log\,N$ stand a chance of having zero entropy (since in that case the few
linear relations defined by~\eqref{eom} could potentially fix the algebra on the remaining
$\sim\log\,N$ elements).
\footnote{Interestingly, the equations of motion for a sprinkling of $O$ in 2$d$ in the local theory 
are such that a large number of points (roughly 50$\%$ and randomly scattered throughout the sprinkling) 
do not appear in the linear relations defined by $A$ at all. So
as far as the dynamics of the field is concerned these points effectively behave as if they
weren't part of the causet. Surprisingly though, going to a cylinder topology seems 
to get rid of these points, yet many of the peculiar results discussed so far continue to hold. 
Therefore it is unclear whether there exists a connection between the infinite entropy and 
the existence of such points.}
To further exacerbate the problem, when the entropy of a subregion is non-zero it is (typically) infinite.

When viewed this way a fix to this infinite entropy presents itself. That is
to increase $k$ so as to ameliorate the equations of motion. In practice this step can be achieved by a deformation of $\Delta$ that sets $\alpha \rightarrow 0$ whenever 
$|\alpha|\le\kappa$, for some $\kappa>0$, and then redefining the vacuum two-point 
function 
\be
W_\kappa:=\text{Pos}(\idelta_\kappa).
\ee
The effect of this deformation of $\Delta$ and $W$, is two-fold. 

First, $\Delta_\kappa$ 
now has a much larger kernel that grows linearly in $N$,
\be
k_\kappa:=\text{dim}(\kerd_\kappa)\approx\beta(\kappa)N,
\ee 
\begin{figure}[h!]
\centering
\includegraphics[scale=0.4]{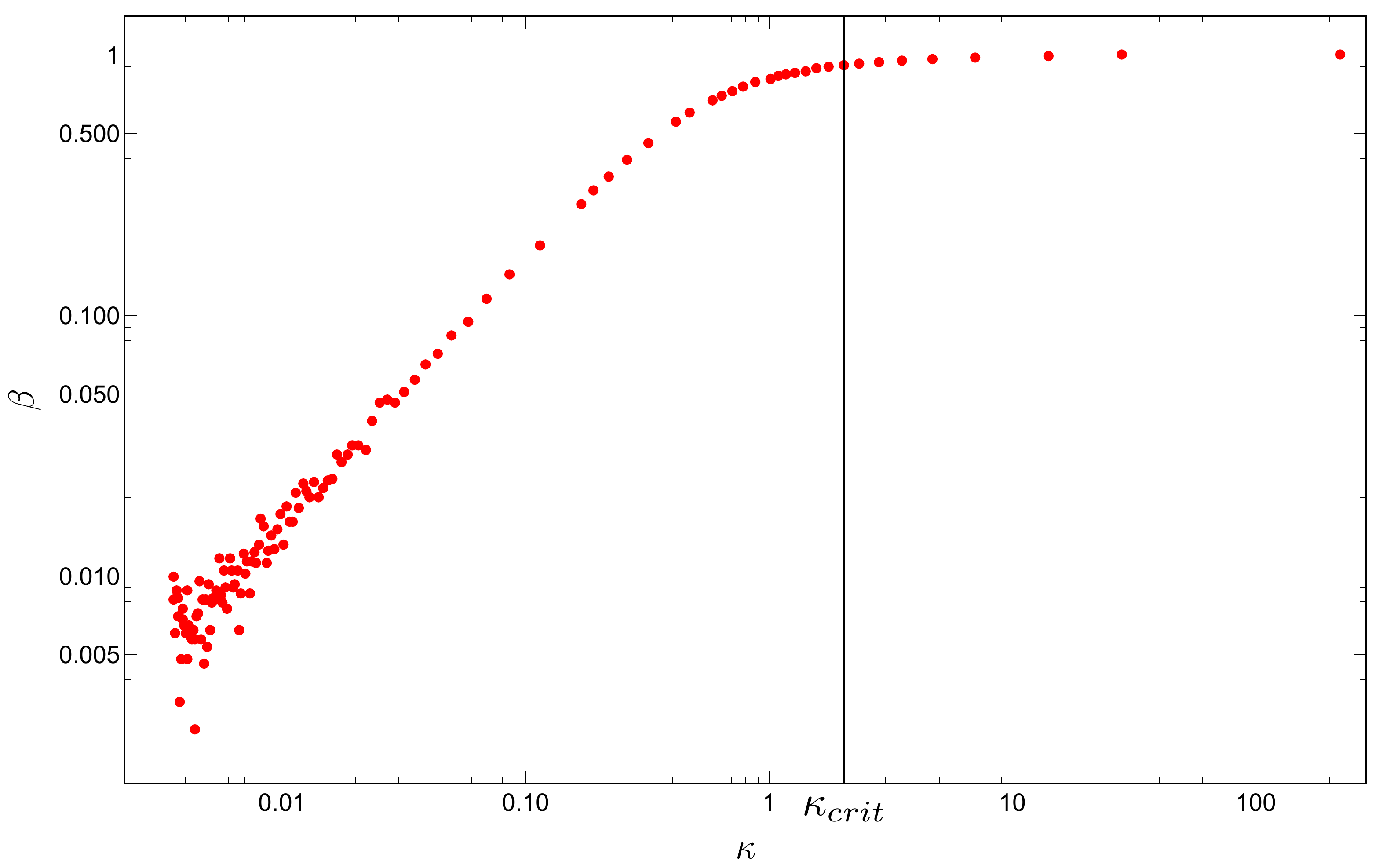}
\caption{A plot of $\beta$ vs. $\kappa$. Note that for large $\kappa>\kappa_{crit}$ 
the value of $\beta$ is approximately constant and close to $1$. As $\kappa$ decreases
below $\kappa_{crit}$ the value of $\beta$ starts decreasing rapidly. This implies that
for values of $0<\kappa\ll\kappa_{crit}$ valid Cauchy regions must be large, while for 
$\kappa_{crit}\lesssim\kappa<1$ they can be relatively small. Note also that $\kappa_{crit}$
corresponds roughly to the point in the spectrum of $\idelta$ just before the kink.}\label{beta}
\end{figure}
for some $\kappa$ dependent coefficient $\beta$, see Figure \ref{beta}.\footnote{Note that for any 
$\kappa > 0$ there exists some $n > 0$ such that for $N>n$,
$\beta(\kappa) > 0$.} 
Hence, given a Cauchy region $\cau$ of size $n$, whose complement 
has size $m=N-n = (1-\gamma)N$, $\gamma=n/N$,
and a cutoff $\kappa>0$, if $\gamma >1- \beta(\kappa)$, $\cau$ stands a chance of having 
zero entropy. Notice that the larger $\kappa$ is, the smaller the Cauchy regions can be while still 
having zero entropy.

Secondly, $\Delta_\kappa$ is now (typically) a fully populated antisymmetric matrix (provided 
the cutoff sets at least one pair of non-zero eigenvalues of $\Delta$ to zero which,
for fixed $\kappa>0$, is guaranteed to be true for sufficiently large $N$), and since
$\idelta_\kappa = [\phi,\phi]$ this means we now have violations of microcausality.
Therefore, by introducing a cutoff on $\text{spec}(\Delta)$ 
we have traded violations of microcausality 
for ``better" equations of motions. 

One can think of the violations of microcausality as 
having effectively turned the causet's partial order into a total order, as far as the quantum
field is concerned. This can also be directly seen by looking at the 
``improved" equations of motion of the field which now fail to respect
the causal structure of the underlying causal set.

This property of $\Delta_\kappa$ implies that the algebra of {\it any} subregion of the 
causal set now has trivial commutant. In the continuum microcausality implies that
only Cauchy regions have trivial commutants, and their entropy is zero if the state is pure.
Translating this to the causal set would therefore suggest that, since the commutant
of any subregion is trivial, every subregion effectively acts like a Cauchy region, and its entropy 
should therefore vanish. 

A caveat to this argument is that the entropy of these subregions 
will be zero only if the subregion is sufficiently large. This is because the equations of motion
are given by a finite number of linear relations \eqref{eom} so that one would not expect
a subregion to be able to ``support" enough initial data if its size is smaller than $N-\beta(\kappa)N$. 

Our numerical simulations do indeed show that the entropy of $W_\kappa$ restricted to 
any sufficiently large subset of the causet is zero. If the subset is made smaller than $N-\beta(\kappa)N$ 
then the entropy becomes non-zero and, in fact,
ill-defined/infinite for large enough $N$, where again we find that $\kerd_\kappa\ne\kerr_\kappa$.

What should one do about the small subregions for which the entropy is non-zero even after the deformation?
Since numerical simulations show that for large enough $N$, their entropy is again infinite because 
$\kerd_\kappa\ne\kerr_\kappa$, a natural fix would be to introduce a secondary cutoff
inside the subregion, like the one discussed in Section~\ref{loccut}.
\footnote{Another approach could be to increase $\kappa$ in the full spacetime, but
while this might ensure that $\kerd_\kappa=\kerr_\kappa$
for some of the smaller regions for which the equality was initially violated, there will 
always be even smaller subregions for which the equality is still violated. Thus, it appears that in order
for arbitrarily small but finite subregions to have a well-defined entropy one also
needs to implement a second cutoff within the subregions.}
This is exactly what we did in Section~\ref{sec:SE}, except that we did not differentiate
between large or small regions (large or small relative to our choice of cutoff), and 
implemented a second cutoff in the subregion irrespective of its size. 
This choice is the consistent one to make if ultimately one wants to think of the subregions
as embedded into full Minkowski space.

An obvious question is whether a second cutoff should also be introduced for Cauchy
regions, since there too there will exist small regions for which $\kerd_\kappa\ne\kerr_\kappa$. 
The answer here seems to be no,
both if one considers the analogous computation in the continuum, where a truncation of $W$
inside Cauchy regions is not needed in order to get a zero entropy, and  the fact that 
introducing a second cutoff again leads to a non-zero entropy that diverges in the limit $\ell\rightarrow0$. 
So, in this case, the fact that a small Cauchy region has non-zero entropy simply means that
it is too small to support the initial data necessary for the equations of motion to be solved, and is 
therefore not a problem that needs fixing.

\subsection{Recap}

We can summarise the results of this section as follows. The SJ-vacuum $W$, when restricted
to proper subregions of the causal set, has infinite entropy. This divergence arises from
purely classical components of $W$, and can interpreted as coming from the nontrivial
centre of the operator algebra associated to the subregion. 

The types of subregions considered can be categorised into two classes: 
Cauchy subregions, and non-Cauchy subregions. 
By comparison with the continuum we expect the entropy of (sufficiently large) members of the former class 
to be zero, while the entropy of the latter should be non-zero. 
One way to ensure that both these expectations are met 
is to deform the global Wightman function into $W_\kappa$, and to implement a secondary cutoff
for any subregion that is not a Cauchy region. If $\kappa$ is chosen such that in both the full causet
and the subcauset one eliminates the part of the spectrum that grossly deviates from the
expected continuum behvaiour, and scales like $N^{1/d}$ in every dimension, 
then an area law is recovered for subregions for which the area law is expected, and the entropy of 
all sufficiently large Cauchy regions vanishes.

\section{Summary and Discussion}
\label{secdisc}

We have numerically studied the entanglement entropy of subregions of sprinklings of causal diamonds in 2, 3 and 4 
dimensional Minkowski spacetime. The quantum field theories for which the entropy was computed were constructed
using the Sorkin--Johnston prescription from both local Green functions and nonlocal Green functions parametrised 
by a nonlocality scale $l_k$. We provided strong numerical evidence that the entanglement entropy for a subdiamond
inside a larger sprinkled diamond in 2$d$, with cutoffs that truncate the part of the spectrum 
of $\idelta$ and $W$ that grossly deviates from the continuum's behaviour both in the larger and smaller diamonds, 
satisfies the usual area law for both classes of theories with a coefficient
that is consistent with the known value in the continuum. 

In higher dimensions and for the same setup we provided preliminary evidence that the local theories are consistent with 
the usual area-law in their respective dimensions, again when a double cutoff is implemented. For the nonlocal theories 
our numerical results suggest that the divergence of the entropy is weaker than  
for the local theories. We argued that this weaker divergence is in accordance with the entanglement entropy of the 
continuum limit of these nonlocal field theories also diverges less strongly as the cutoff scale is sent to zero. 
The preliminary numerical results in the causal set are consistent with these continuum calculations.

Having established this correspondence with the area law in the continuum when a double cutoff is introduced, we
studied the effect of the two cutoffs taken independently on two classes of subregions: 
Cauchy regions and (globally hyperbolic) non-Cauchy regions. We provided evidence that in the absence of any cutoff 
the entropy of (almost --- see discussion after Eq.~\eqref{eq:A}) any proper subset of the
causal set is infinite due to classical components of the Wightman function that arise whenever $\kerr \subset \kerd$. 
These infinite contributions can be 
ascribed to the Shannon entropy associated to the non-trivial centre of the operator algebra of the subregion, which
is known to be infinite in the case of normally distributed random variables.

We argued that a natural way to eliminate this divergence is to augment $\kerr$, so that it matches
$\kerd$, by implementing a local cutoff. Imposing this matching condition trivialises the centre of the algebra by ensuring 
that an irreducible representation of the algebra exists in which the non-trivial elements of the centre are identically
zero. Having implemented this condition we found that
the remaining entropy, which is now an entanglement entropy, follows a spacetime volume law, rather
than an area law, in all dimensions considered and for any subregion. We speculate that the same is true in any 
dimensions for both local and nonlocal theories. 

By further increasing the local cutoff (thus augmenting $\kerr$ and $\kerd$ while 
preserving the matching condition), we found that the divergence of the entanglement entropy with the discreteness 
scale weakens. In particular, in 2$d$ at least, if one forces the part of the spectrum that grossly deviates from 
the continuum's spectrum into the kernels of $R$ and $\Delta$, then an area law is recovered (a logarithmic dependence), but with the wrong coefficient.
While this may be a welcome result for globally hyperbolic subregions that are not Cauchy regions, it is not so 
for Cauchy regions for which one would expect the entropy to be zero. 

This major discrepancy between the entropy of Cauchy regions in the causal set and in the continuum appears to stem
from poorly defined equations of motion on the causal set which, in turn, is a consequence of the smallness of $\kerd$ 
on the causal set. Implementing a global cutoff on $\Delta$ that augments its kernel before 
defining the SJ-vacuum, enhances the equations of motion at
the expense of introducing violations of microcausality. 
The effect of this enhancement introduced by the global cutoff is that now sufficiently large Cauchy regions have zero entropy.
While the effect of the local cutoff is to ensure that the microcausality violations do not spoil the finiteness and area law of the entanglement entropy 
of non-Cauchy subregions, irrespectively of their size.

These results, while shedding some light on the role played by the global and local cutoffs, still leave
many questions unanswered. The most pressing of is the nature of the SJ vacuum, with and without
the global cutoff. 

Notice first that the global cutoff is not strictly needed if all one wants is to make the entropy
of subregions well defined, a local cutoff will suffice. In its absence however, one must confront
the fact that the algebras associated to subregions of the causal set have non-trivial centres,
something of relevance even outside the context of entanglement entropy. 
This inordinate degree of nonlocality appears to stem from poorly defined equations
of motion but, as we have shown, can be contained by introducing a global cutoff that
enhances the equations of motion. 
The resulting (global) operator algebra can now be meaningfully restricted to subregions, 
in the sense that the algebras of the subregions admit irreducible representations, but 
at the expense of having introduced violations of microcausality.
What's remarkable is that in the resulting theory the entropy
of a subdiamond, when taken in conjunction with a suitable choice of local cutoff, 
reproduces the continuum result. 

Assuming that all of our observations continue to hold as $N\rightarrow\infty$,
while holding the discreteness scale fixed, e.g. in the case of a sprinkling of infinite Minkowski 
spacetime, and noting in particular that $\text{dim(}\kerd)\rightarrow\infty$ in this limit,
we can argue that a global cutoff will not be necessary. 
To this end note first that in the continuum limit of the 2$d$ example considered in~\cite{Saravani:2013nwa}, 
no global cutoff is required to ensure that the entropy Cauchy regions vanishes.\footnote{While 
this hasn't been checked explicitly, it should hold by construction.} This is ultimately
because the equations of motion establish an equivalence between the algebras of Cauchy
regions (including Cauchy surfaces) and the full algebra. Now in the infinite causal set
a similar argument can be made with the Klein-Gordon equations of the continuum 
replaced by their discrete counterpart which, being infinite in number, should  
guarantee an equivalence between the global algebra and the algebras of Cauchy regions 
(at least for ``thick" Cauchy regions, if not for maximal antichains).

If true, this argument would imply that the global cutoff introduced in this paper and
in~\cite{Sorkin:2016pbz} is only needed for finite size causal sets.
Note however that when restricting the state defined on an infinite causet to a finite size subregion we see no 
a priori reason why the matching condition should still be satisfied. If it is not, then a local cutoff would once
again be required in order to make the entropy finite. Also, it seems plausible that within a finite region
the spectrum of $\Delta$ will once again possess a part of the spectrum that deviates from the continuum's,
in which case the spacetime volume law would likely endure.
Be that as it may, the peculiar features of the Sorkin-Johnston vacuum for finite size causal sets exist and call for a 
more thorough investigation

To conclude, it is interesting that the entropy of the SJ vacuum on the causal set is infinite despite the underlying
discreteness. Having argued that it is ultimately a consequence of the global nature of the definition of the SJ vacuum,
together with its poorly defined equations of motion, it is reasonable to expect that this will be a feature
of SJ vacua on generic discrete spacetimes (indeed we have results that confirm this in the case of regular lattice
discretisations of the 2$d$ diamond). This begs the question of whether there exist vacua on discrete spacetimes, 
other than the SJ-vacuum defined from microcausality violating Pauli-Jordan functions, whose restriction to subregions
directly leads to a well-defined, finite entanglement entropy. Given that the emergence of an area law seems to be tied with the imposition of the two cutoffs 
studied here, one might wonder if and how this law would be preserved should these vacua exist.
Either way, these findings will surely have relevant implications for the role and meaning of entanglement entropy 
in black hole spacetimes.

\section*{Acknowledgments}

The authors would like to thank Yasaman Yazdi for useful discussions. AB  wish  to  acknowledge  the  support  of the Austrian Academy of Sciences through Innovationsfonds  ”Forschung,  Wissenschaft  und  Gesellschaft“,  and
the University of Vienna through the research platform TURIS. This publication was made possible through the support of the grant from the John Templeton Foundation No.51876. The opinions expressed in this publication are those of the authors and do not necessarily reflect the views of the John Templeton Foundation.

      \bibliographystyle{plain}
\bibliography{sentropy}
			
\begin{appendices}

\section{Entanglement entropy of nonlocal scalar fields via the replica trick}\label{appA}

In this appendix, we briefly review the computation of the entanglement entropy of a quantum field using the \textit{replica trick} \cite{Callan:1994py,Nesterov:2010yi} and use it in the case of the nonlocal scalar field theory emerging from CST in the continuum limit in two, three and four spacetime dimensions.

\subsection{The replica trick}

Let us consider a quantum field $\phi (x)$ on a $d$-dimensional spacetime with coordinates $x^\mu=(\tau,x,z^i,i=1,...,d-2)$, where $\tau$ is the Euclidean time, and a hypersurface $\Sigma$ defined by the condition $x=0$. The coordinates $z^i$ are therefore the coordinates on $\Sigma$.  

Entanglement entropy is computed by preparing the field in the vacuum state and then tracing out the degrees of freedom which are inside (outside) the surface $\Sigma$. The computation goes as follows.

First, we define the vacuum state of scalar field by a path integral over half of the Euclidean space defined by $\tau\leq0$ in such a way that the field assumes the boundary condition $\phi (\tau=0,x,z)=\phi_0(x,z)$,
\begin{equation}
\Phi [\phi_0 (x,z)]=\int\displaylimits_{\phi(x^\mu)|_{\tau=0}=\phi_0 (x,z)} \mathcal{D} \phi\;e^{-W[\phi]},
\end{equation}
where $W[\phi]$ is the action of the field. The surface $\Sigma$, given by $(\tau=0,x=0)$, separates the boundary data in two parts $\phi_- (x,z)$ for $x<0$ and $\phi_+ (x,z)$ for $x>0$. Now tracing over $\phi_-(x,z)$ one obtains a reduced density matrix
\begin{equation}\label{density_m}
\rho (\phi_+^1,\phi_+^2)=\int \mathcal{D}{\phi_-}\; \Phi(\phi_+^1,\phi_-)\Phi(\phi_+^2,\phi_-),
\end{equation}
where the path integral goes over fields defined on the whole Euclidean space-time except a cut $(\tau= 0,x >0)$. In the path integral the field $\phi(x^\mu)$ takes a boundary value $\phi^2_+$ above the cut and $\phi^1_+$ below the cut.

The trace of $n$-th power of the density matrix (\ref{density_m}) is given by the Euclidean path integral over fields defined on an $n$-sheeted covering of the cut space-time. Essentially one considers $n$ copies of this space-time attaching one copy to the next through the cut gluing analytically the fields. Passing from Cartesian coordinates $(\tau,x)$ to polar ones $(r,\alpha)$, the cut corresponds to the values $\alpha=2 \pi k$ with $k=1,2,...,n$. This $n$-fold space is geometrically a flat cone $C_n$ with a deficit angle $2\pi (n-1)$. Therefore one has
\begin{equation}\label{conical_path}
\text{Tr} \rho^n=Z[C_n],
\end{equation}
where $Z[C_n]$ is the Euclidean path integral over the
n-fold cover of the Euclidean space, i.e. over the cone $C_n$.

It can be shown that it is possible in (\ref{conical_path}) to analytically continue to non-integer values of $n\rightarrow \beta$. With that said, one observes that $-\text{Tr} \hat{\rho} \ln \hat{\rho}=-(\beta \partial_\beta -1) \ln \rho^\beta |_{\beta=1}$, where $\hat{\rho}=\frac{\rho}{\text{Tr} \rho}$. In polar coordinates $(r,\alpha)$, the conical space $C_\beta$ is defined by making the coordinate $\alpha$ periodic with the period $2\pi\beta$, where $(1-\beta)$ is very small. Then introducing $W(\beta)=\ln Z[C_\beta]$, one has
\begin{equation}
S=(\beta\partial_{\beta}-1)W(\beta)|_{\beta=1}.
\end{equation}
At this point in order to calculate $W(\beta)$ one can use the heat kernel method in the context of manifolds with conical singularities (see \cite{Nesterov:2010yi} and references therein).

Once the effective action $W(\beta)$ is calculated, the entanglement entropy is simply given by the following formula
\begin{equation}\label{ent_entropy}
S=\frac{A(\Sigma)}{12 (4\pi)^{(d-2)/2}}\int_{\epsilon^2}^\infty \frac{ds}{s}\; P_{d-2}(s), 
\end{equation}
where $\epsilon$ is a UV cut-off that makes the integral finite ($s$ has dimensions of a length squared) and
\begin{equation}\label{coneheatk}
P_{d-2}(s)=\frac{2}{\Gamma(\frac{d-2}{2})}\int_0^\infty dp \;p^{d-3}\;e^{-sF(p^2)}.
\end{equation}

$F(p^2)$ is the Fourier transform of kinetic operator of a non-interacting Lorentz invariant scalar field theory.

\subsection{The case of the nonlocal scalar field theories from CST}

We will now apply the procedure described above to the case of nonlocal scalar field theories from CST. In particular we will consider the continuum d'Alembertians obtained by averaging the operators \eqref{nonlocalop} over all sprinklings of Minkowski spacetime. The result of the averaging process is given by the following expression
\begin{equation}\label{contnlop}
\Box^{(d)}_{\rho_k} \phi(x)=\rho^{2/d}\left(a\phi(x)+\rho\sum_{n=0}^{L_{\rm max}}\frac{b_n}{n!}\int\displaylimits_{J^-(x)}e^{-\rho V(x,y)}\,[\rho V(x,y)]^n\,\phi(y)\,dy\right),
\end{equation}
where $\rho_{k}=l_k^{-d}$, $l_{k}$ being the nonlocality scale, $J^-(x)$ is the causal past of $x$ and $V(x,y)$ is the spacetime volume between the past light cone of $x$ and the future light cone of $y$.

Following the discussion in \cite{Aslanbeigi:2014zva}, the momentum space representation of \eqref{contnlop} can be considered for $p^2\ll\rho_k^{-2/d}$ (IR limit) and $p^2\gg\rho_k^{-2/d}$ (UV limit). The former is universal and given by
\begin{equation}\label{FIR}
F_{\rho_k}(p^2)\rightarrow -p^2,\quad \text{for}\quad p^2\ll\rho_k^{-2/d},
\end{equation}
while the latter depends on the spacetime dimensions and can be given as
\begin{equation}\label{FUV}
F_{\rho_k}(p^2)\rightarrow a\,\rho_k^{2/d}+b\,\rho_k^{2/d+1}k^{-d}+...,\quad\text{for}\quad p^2\gg\rho_k^{-2/d}.
\end{equation}

From Eq.\eqref{FUV}, one can see that the nonlocal d'Alembertian goes to a constant in the UV. This term correspons to a delta function for the Green functions in real space in the coincident limit and it is essentially a remnant of the fundamental discreteness of the causal set (see the discussion in \cite{Aslanbeigi:2014zva}). This term can be subtracted and one can define a regularized d'Alembertian operator as
\begin{equation}\label{regop}
F_\textrm{reg}(p^2)=\frac{a\,\rho_{k}^{2/d} F_{\rho_k}(p^2)}{a\,\rho_k^{2/d}-F_{\rho_k}(p^2)}.
\end{equation}

The operator \eqref{regop} maintains the correct IR limit given by Eq.\eqref{FIR} and possesses the new UV behavior displayed by the following expression
\begin{equation}
F_\textrm{reg}\rightarrow-\frac{a^2}{b}\rho_k^{2/d-1}\,p^d+...,\quad\text{for}\quad p^2\gg\rho_k^{-2/d}.
\end{equation} 
In order to compute the entanglement entropy via the replica trick we need to Wick rotate the operator $F_\textrm{reg}$ or, equivalently, its retarded propagator. However this cannot be done on the retarded propagator because the contour, $\Gamma_R$, would cross singularities. To avoid this problem one must use the Feynman propagator whose contour can be Wick rotated without crossing any singularities (see \cite{Belenchia:2014fda,Saravani:2015rva,Belenchia:2015aia} for further details).

\subsubsection{IR and UV behavior of the entanglement entropy}

The behavior of \eqref{regop} for $p^2\ll\rho_k^{2/d}$, for negligible nonlocal effects, is given by eq.\eqref{FIR}. Hence the entanglement entropy computed solely on the basis of this contribution scales with the area of the surface $\Sigma$. In particular, in $d=2,3,4$, the entropy is given by
\begin{equation}
\begin{split}
& S^{(2)}_{loc}=\frac{1}{6} A(\Sigma) \ln \left(L/\epsilon\right)\\
& S^{(3)}_{loc}=\frac{A(\Sigma)}{12 \sqrt{\pi}}\frac{1}{\epsilon}\\
&S^{(4)}_{loc}=\frac{A(\Sigma)}{48 \pi}\frac{1}{\epsilon^2},
\end{split}
\end{equation}
where $L$ is a IR cutoff and $\epsilon$ is a UV cutoff needed to make the entanglement entropy finite.

In the UV the entropy is dominated by the UV behavior of the momentum space d'Alembertian. For $d=2,3,4$ the expansion of \eqref{regop} in the limit $p^2\gg\rho_k^{2/d}$ is given by the following expressions
\begin{equation}
\begin{split}\label{UVexp}
&F^{(2)}(p^2)\rightarrow-\frac{p^2}{2},\\
&F^{(3)}(p^2)\rightarrow-\frac{p^3}{\rho_k^{1/3}},\\
&F^{(4)}(p^2)\rightarrow-\frac{p^4}{\rho_k^{1/2}}.\\
\end{split}
\end{equation}

By using \eqref{UVexp} in \eqref{ent_entropy} and \eqref{coneheatk}, one can estimate the leading contribution to the entanglement entropy in the limit $p^2\gg\rho_k^{2/d}$ for the nonlocal models in the continuum. The results are 
\begin{equation}
\begin{split}
&S^{(2)}_{UV}\propto A(\Sigma) \ln \left(L/\epsilon\right),\\
&S^{(3)}_{UV}\propto\frac{A(\Sigma)}{\epsilon^{2/3}\,l_k^{1/3}},\\
&S^{(4)}_{UV}\propto\frac{A(\Sigma)}{\epsilon\,l_k}.
\end{split}
\end{equation}

In $d=3,4$, the scaling of the entropy with respect to the cutoff $\epsilon$ is weaker with respect to the local case due to the presence of the nonlocality scale. In $d=2$, the nonlocality scale does not enter the UV expansion of the wave operator, hence the leading contribution to the entanglement entropy in the UV is untouched with respect to the local theory.

\end{appendices}

\end{document}